\documentclass[12pt]{article}
\pdfoutput=1
\usepackage[utf8]{inputenc}
\usepackage[DIV=13]{typearea}
\usepackage{ragged2e}
\usepackage{amsmath,amsfonts,bbm,mathrsfs,amssymb}
\usepackage{slashed,cancel}
\usepackage[usenames,dvipsnames]{xcolor}
\usepackage{cite}
\usepackage{bm} 
\usepackage{hyperref}
\hypersetup{colorlinks=true,urlcolor=Magenta,anchorcolor=blue,citecolor=blue,filecolor=blue,
            linkcolor=Magenta,menucolor=blue, linktocpage=true,pdfproducer=medialab}
\usepackage[english]{babel}
\usepackage{catchfile}
\usepackage{indentfirst}
\usepackage{graphicx}
\usepackage{float}
\usepackage{enumerate}
\usepackage{subcaption}
\usepackage{multirow}
\usepackage{tikz-feynman}
\usepackage[normalem]{ulem} 

\newcommand{\email}[1]{\href{mailto:#1}{\tt #1}}

\numberwithin{equation}{section}

\newcommand{\blue}[1]{\color{blue} #1 \color{black}}

\newcommand{\magenta}[1]{\color{Magenta} #1 \color{black}}
\newcommand{\be}{\begin{equation}}
\newcommand{\ee}{\end{equation}}
\newcommand{\ba} {\begin{equation}\begin{aligned}}
\newcommand{\ea} {\end{aligned}\end{equation}}

\newcommand{\sL}{\mathscr{L}}

\newcommand{\cG}{\mathcal{G}}
\newcommand{\cY}{\mathcal{Y}}
\newcommand{\cH}{\mathcal{H}}
\newcommand{\cO}{\mathcal{O}}
\newcommand{\cM}{\mathcal{M}}
\newcommand{\cN}{\mathcal{N}}
\newcommand{\cR}{\mathcal{R}}

\newcommand{\unity}{\mathbbm{1}}
\newcommand{\derp}{\partial}
\newcommand{\hc}{\text{h.c.}}
\newcommand{\nn}{\nonumber}

\newcommand{\vev}[1]{\langle{#1}\rangle}
\newcommand{\ov}[1]{\overline{#1}}
\newcommand{\sign}[1]{\text{sgn}(#1)}

\newcommand{\meV}{\ \text{meV}}
\newcommand{\eV}{\ \text{eV}}

\newcommand{\TeV}{\ \text{TeV}}
\newcommand{\GeV}{\ \text{GeV}}

\def\diag{{\tt diag}}

\def\BR{{\rm BR}}

\def\vep{\varepsilon}
\def\tH{\widetilde{H}}
\def\UPQ{U(1)_\text{PQ}}

\begin{document} 
\renewcommand*{\thefootnote}{\fnsymbol{footnote}}
\begin{titlepage}

\vspace*{-1cm}
\flushleft{\magenta{IFT-UAM/CSIC-22-34}}
\\[1cm]

\begin{center}
\bf\LARGE \blue{
Dynamical Minimal Flavour Violating Inverse Seesaw}\\[4mm]
\centering
\vskip .3cm
\end{center}
\vskip 0.5  cm
\begin{center}
{\large\bf Fernando Arias-Arag\'on}${}^{a)}$~\footnote{\email{arias@lpsc.in2p3.fr}},
{\large\bf Enrique Fern\'andez Mart\'inez}${}^{b)}$~\footnote{\email{enrique.fernandez-martinez@uam.es}},\\[2mm]
{\large\bf Manuel Gonz\'alez-L\'opez}${}^{b)}$~\footnote{\email{manuel.gonzalezl@uam.es}}, and 
{\large\bf Luca Merlo}${}^{b)}$~\footnote{\email{luca.merlo@uam.es}}
\vskip .7cm
{\footnotesize
${}^{a)}$ Laboratoire de Physique Subatomique et de Cosmologie,\\ 
Université Grenoble-Alpes CNRS/IN2P3, Grenoble INP, 38000, Grenoble, France
\vskip .2cm
${}^{b)}$ Departamento de F\'isica Te\'orica and Instituto de F\'isica Te\'orica UAM/CSIC,\\
Universidad Aut\'onoma de Madrid, Cantoblanco, 28049, Madrid, Spain
}
\end{center}
\vskip 2cm
\begin{abstract}
\justify 

The Inverse Seesaw mechanism is dynamically realised within the Minimal Lepton Flavour Violation context. Lepton number, whose breaking is spontaneously realised, is generalised to a global Abelian factor of the whole flavour symmetry, that also plays the role of the Peccei-Quinn symmetry. The associated Goldstone boson is a Majoraxion that solves the Strong CP problem and represents a Dark Matter candidate.

Three distinct scenarios are identified in terms of flavour symmetry and transformation properties of the exotic neutral leptons that enrich the Standard Model spectrum. The associated phenomenology is studied, focusing on the deviations from unitarity of the PMNS mixing matrix. The strongest constraints arise from the determination of the number of active neutrinos through the invisible width of the $Z$, the comparison of the measured $W$ boson mass with its prediction in terms of the Fermi constant from muon decay, and the null searches for the radiative rare muon decay and $\mu\to e$ conversion in nuclei. The heavy neutral leptons may have masses of a few TeV, leaving open the possibility for a direct detection at future colliders.

The impact of the recent measurement of the $W$ mass at the CDF II detector has also been considered, which, in one of the scenarios, points to a sharp prediction for the masses of the heavy neutral leptons at about $2-3$ TeV.

\end{abstract}
\end{titlepage}
\setcounter{footnote}{0}

\pdfbookmark[1]{Table of Contents}{tableofcontents}
\tableofcontents

\renewcommand*{\thefootnote}{\arabic{footnote}}

\section{Introduction}
\label{sec:intro}

Neutrino oscillations and their interpretation in terms of massive neutrinos represent the first historical evidence of beyond the Standard Model (SM) physics. While only one effective operator can be written at low energies to describe contributions to neutrino masses, that is the Weinberg operator~\cite{Weinberg:1979sa}, several ultraviolet completions have been proposed. Among them, low-scale variants of the type-I Seesaw model, such as the so-called Inverse Seesaw  (ISS) mechanism~\cite{Wyler:1982dd,Mohapatra:1986bd,Bernabeu:1987gr}, are particularly appealing. In these Seesaw variants, the smallness of neutrino masses, or equivalently of the Weinberg operator, is tied to an approximate lepton number symmetry. Conversely, the Seesaw scale can be comparatively low and the mixing of the new states with the active neutrinos, encoded by the only $d=6$ operator that is obtained at tree level~\cite{Broncano:2002rw}, large enough to lead to interesting and testable phenomenology. Thus, these setups may be testable both at colliders, focusing on signals from the exotic neutral leptons, as well as in precision electroweak and flavour observables.

The ISS is a specific realisation of a wider class of Low-Scale Seesaw (LSS) frameworks~\cite{Kersten:2007vk,Abada:2007ux}, that follow from introducing two types of exotic neutral leptons (which will be labelled in the following as $N_R$ and $S_R$), with opposite transformation properties under the lepton number symmetry, which is approximately enforced.  Using a compact notation for all of the neutral leptons,
\be
\chi\equiv(\nu_L,\, N_R^c,\, S_R^c)^T\,,
\ee
being $\nu_L$ the neutral component of the EW lepton doublet $\ell_L$, the LSS characteristic mass term reads
\be
-\sL_Y\supset\dfrac{1}{2}\ov{\chi}\cM_\chi\chi^c+\hc\,,
\label{GenericNeutralMassLag}
\ee
where
\be
\cM_\chi=
\begin{pmatrix}
0 & \dfrac{v}{\sqrt2} \cY_\nu & \epsilon\dfrac{v}{\sqrt2} \cY'_\nu\\
\dfrac{v}{\sqrt2} \cY_\nu^T & \mu' & \Lambda\\[4mm]
\epsilon\dfrac{v}{\sqrt2} \cY^{\prime T}_\nu & \Lambda^T & \mu \\
\end{pmatrix}.
\label{genericISSmatrix}
\ee
Here, $v=246\GeV$ is the vacuum expectation value (VEV) of the Higgs doublet $H$, $\cY_\nu$ and $\cY'_\nu$ are the Dirac Yukawa matrices that couple the EW lepton doublet to the exotic neutral leptons $N_R$ and $S_R$ respectively, $\Lambda$, $\mu$ and $\mu'$ are matrices in the flavour space of $N_R$ and $S_R$, and finally $\epsilon$ is a real parameter.

Under the assumption of an approximate lepton number symmetry, $\mu$, $\mu'$ and $\epsilon$ (or their elements) must be small parameters, while $\Lambda$ dominates over the other entries. The light active neutrinos develop a suppressed mass that has two tree level contributions,
\be
m_\nu\simeq\dfrac{v^2}{2}\left[\left(\cY_\nu\dfrac{1}{\Lambda^T}\mu\dfrac{1}{\Lambda}\cY_\nu^T\right)-\epsilon\left(\cY'_{\nu}\dfrac{1}{\Lambda}\cY_\nu^T+\cY_{\nu}\dfrac{1}{\Lambda^T}\cY^{\prime T}_\nu\right)\right]\,.
\label{genericISSmnu}
\ee
A back-of-the-envelope calculation reveals that to obtain the minimum value required for the neutrino mass to reproduce the observed atmospheric mass splitting of $0.05\eV^{2}$, with $\Lambda$ of $\cO(\TeV)$ and $v/\Lambda\sim0.01$ in order to comply with present constraints, then $\mu\sim$ KeV and $\epsilon\sim10^{-10}$. This is the origin of the most interesting feature of LSS mechanisms: while neutrino masses remain small, on one hand, the heavy neutral leptons are relatively light and possibly detectable at colliders; and on the other hand, integrating these states out, a unique $d=6$ operator is generated at tree level~\cite{Broncano:2002rw}:
\be
\cO_{d=6}\simeq \ov{\ell_L}\tH\, c_{\cO_{d=6}}\,i\,\slashed{\derp}\left(\tH^\dag\ell_L\right)\,
\qquad\text{with}\qquad
c_{\cO_{d=6}}=\cY_\nu\dfrac{1}{\Lambda\Lambda^\dag}\cY^\dag_\nu\,,
\label{Genericd6Operator}
\ee
where $\tH\equiv i\sigma_2H^\ast$. This operator does not depend on neither $\mu^{(\prime)}$ nor $\epsilon$; thus, its  Wilson coefficient is not suppressed, leading to possibly interesting phenomenological effects in direct and indirect searches.

As shown in Ref.~\cite{Gavela:2009cd}, in the case with only one $N_R$ and one $S_R$ it is possible to completely determine $\cY_\nu$ and $\cY_\nu'$, and therefore the Wilson coefficient of $\cO_{d=6}$ can be uniquely linked to the active neutrino masses. This represents a very predictive scenario: the radiative rare charged lepton decay rates can be expressed in terms of active neutrino masses, lepton mixing angles and the Majorana phases. This is in general not true if more exotic neutral leptons are present in the spectrum. 

Nevertheless, predictivity may be recovered through a flavour symmetry: we will focus in the Minimal Flavour Violating (MFV) scenario~\cite{Chivukula:1987py,DAmbrosio:2002vsn}. The fundamental idea of the MFV setup is that the only source of flavour and CP violation in any physics beyond the SM (BSM) is represented by the SM Yukawa couplings. Technically, this can be obtained by imposing in the whole Lagrangian of the model the flavour symmetry of the kinetic terms, which is a $U(3)$ factor for each fermion species in the spectrum. Yukawa terms can be made invariant under it by promoting the Yukawa matrices to spurion fields, that transform under the flavour symmetry. Masses and mixing then arise once the Yukawa spurions acquire a specific background value. This approach has been first implemented in the quark sector and only subsequently extended to leptons, MLFV~\cite{Cirigliano:2005ck,Davidson:2006bd,Gavela:2009cd,Alonso:2011jd}, although additional hypotheses are necessary in the latter case in order to achieve the same level of predictive power. Indeed, while in the quark sector there are only two Yukawa spurions, in the leptonic one there are typically at least three. To consider a specific BSM realisation, in the type-I Seesaw~\cite{Minkowski:1977sc,Gell-Mann:1979vob,Yanagida:1979as,Mohapatra:1979ia}, besides the charged lepton Yukawa, the neutrino Dirac Yukawa coupling and the exotic neutral Majorana mass should be promoted to spurions in order to enforce the flavour symmetry invariance in the Yukawa Lagrangian. As a result, the active neutrino mass matrix and the $d=6$ operator Wilson coefficients depend on different combinations of the neutrino spurions, preventing in this way a direct link between the different observables. To circumvent this difficulty, two solutions in terms of symmetries have been proposed in the context of the type-I Seesaw mechanism. The first is to reduce the $U(n)$ associated to the $n$ exotic neutral leptons to $O(n)$, with the consequence of having a Majorana mass matrix proportional to the identity. The second, that can only be viable if $n=3$, consists in identifying the $U(3)$ factors of the EW lepton doublet and of the exotic neutral leptons in a common $U(3)$ term, and as a result it is the neutrino Dirac Yukawa matrix the one that is proportional to the identity. In both cases, only one neutrino spurion is left in the Lagrangian, recovering in this way the strong predictive power that characterises MFV in the quark sector.

The MFV ansatz helps providing an efficient control over the new physics (NP) contributions to a large variety of flavour observables. In an effective field theory approach, the Wilson coefficients of higher dimensional operators are made invariant under the flavour symmetry by inserting powers of the spurions, providing suppressions in terms of fermion masses and mixing.~\footnote{The top Yukawa is an exception, as it cannot be considered as an expansion parameter. This aspect has been discussed in Ref.~\cite{Kagan:2009bn}, proposing a resummation procedure for the up-type quark Yukawa matrix.} The result is that, while the bounds on the NP scale in the absence of any flavour symmetry would be of hundreds of TeVs~\cite{Isidori:2010kg,EuropeanStrategyforParticlePhysicsPreparatoryGroup:2019qin}, in the M(L)FV case the NP scale can be at the TeV scale as proven in several contexts~\cite{Chivukula:1987py,DAmbrosio:2002vsn,Cirigliano:2005ck,Davidson:2006bd,Gavela:2009cd,Alonso:2011jd,Cirigliano:2006su,Grinstein:2006cg,Paradisi:2009ey,Grinstein:2010ve,Feldmann:2010yp,Guadagnoli:2011id,Buras:2011zb,Buras:2011wi,Alonso:2012jc,Alonso:2012pz,Lopez-Honorez:2013wla,Barbieri:2014tja,Alonso:2016onw,Crivellin:2016ejn,Dinh:2017smk,Merlo:2018rin,Arias-Aragon:2020qip,Alonso-Gonzalez:2021jsa,Alonso-Gonzalez:2021tpo}. Despite this success, M(L)FV is not a flavour model, as masses and mixing can only be described and not predicted: in other words, the background values of the spurions are simply assumed as working hypotheses. A few attempts to promote M(L)FV to a proper model have been proposed in Refs.~\cite{Feldmann:2009dc,Alonso:2011yg,Nardi:2011st,Alonso:2012fy,Alonso:2013mca,Alonso:2013nca}, where the spurions have been promoted to dynamical scalar fields and the minima of the corresponding scalar potential have been studied. Although interesting results have been found, a definitive answer is still missing (advances in this respect have been done going beyond the MFV ansatz~\cite{Arias-Aragon:2020bzy}), and for this reason we will adopt the traditional spurion formulation in what follows. 

The implementation of MLFV in the ISS scenario has partially been discussed in previous literature, such as in Refs.~\cite{Gavela:2009cd,Dolan:2018yqy}. We revisit and complete the study of MLFV in the ISS context, identifying three different minimal and predictive setups and analysing their phenomenological impact. 

In contrast to the traditional construction of the ISS, we provide a dynamical explanation of the smallness of the $\mu^{(\prime)}$ and $\epsilon$ parameters, by enforcing the invariance of the Lagrangian under an additional $U(1)$ global symmetry. This is an Abelian subgroup of the flavour symmetry of the kinetic terms. Although the idea of a dynamical origin for these small parameters is not new, see for example Refs.~\cite{Ma:2009gu,Bazzocchi:2010dt,DeRomeri:2017oxa,Mandal:2021acg,Fernandez-Martinez:2021ypo}, it is the first time that it is linked to the flavour problem. In particular, we argue that, in order to explain the smallness of $\mu^{(\prime)}$ and $\epsilon$, instead of the usual choice of lepton number, a larger symmetry can be considered, providing further links to the quark sector. We will identify this $U(1)$ term with the Peccei-Quinn (PQ) symmetry~\cite{Peccei:1977hh} and discuss its consequences. 

The spontaneous breaking of the additional $\UPQ$ leads to the appearance of a Goldstone boson (GB), either an axion or an axion-like-particle (ALP). In the first case, the traditional solution to the Strong CP-problem~\cite{Peccei:1977hh,Wilczek:1977pj,Weinberg:1977ma} can be achieved in this context, without introducing any additional ingredient beyond the ones we have already described. The presence of an axion, and of an ALP, in the MFV context has already been presented in Ref.~\cite{Arias-Aragon:2017eww}, discussing in detail only the quark sector. In this paper, we extend the results of the MFV axion (MFVA) model to the lepton sector, identifying the axion with a Majoron, sometimes called Majoraxion in the literature. Last but not least, for a large region of the parameter space we are interested in, the Majoraxion represents a valid Dark Matter (DM) candidate. 

In summary, the model presented in this paper provides a dynamical realisation of the ISS mechanism, consistent with the M(L)FV ansatz to guarantee the necessary flavour protection and thus avoiding unobserved signals at experiments, while providing at the same time a natural solution to the Strong CP problem and a valid DM candidate.

The paper is structured as indicated in the Table of Contents, with a summary of the MLFV setup and the identification of the three dynamical ISS realisations in Sect.~\ref{sec:MFV}. Their technical details and phenomenological analyses are presented in Sects.~\ref{sec:CASEA}, \ref{sec:CASEB} and \ref{sec:CASEC}. Sect.~\ref{sec:MWnew} is devoted to analyse the impact of the recent measurement of the $W$ mass at the CDF II detector~\cite{CDF:2022hxs}. Concluding remarks are in Sect.~\ref{sec:Concls}.

\section{Dynamical Inverse Seesaw within MLFV}
\label{sec:MFV}

Except otherwise stated, we will consider the SM spectrum with the addition of six exotic neutral leptons, divided into two groups, three $N_R$ and three $S_R$. We will see later in this section that the main difference, but not the only one, between these two types of fermions resides in their transformations properties under $\UPQ$. 

The largest possible symmetry of the kinetic terms of the Lagrangian in the lepton sector is 
\be
\cG_F=U(3)_{\ell_L}\times U(3)_{e_R}\times U(3)_{N_R}\times U(3)_{S_R}\,,
\ee
where each term corresponds to a lepton species. $\cG_F$ can be rewritten as a product of the non-Abelian terms,
\be
\cG_F^\text{NA}=SU(3)_{\ell_L}\times SU(3)_{e_R}\times SU(3)_{N_R}\times SU(3)_{S_R}\,,
\ee
under which the corresponding leptons transform as triplets, and of four Abelian factors that can be recasted to make explicit some interesting properties of the theory. For a complete discussion we need to include the three Abelian factors of the flavour symmetry of the quark kinetic terms. The product of the seven $U(1)$ symmetries can be rewritten as the product of the global version of the hypercharge $U(1)_Y$, baryon number $B$, lepton number $L$, the PQ symmetry $\UPQ$, and three rotations on $e_R$, $N_R$ and $S_R$. 

The non-Abelian symmetry content $\cG_F^\text{NA}$ is the one responsible for the description of the lepton mass hierarchies and mixing~\cite{Feldmann:2009dc,Alonso:2011yg,Nardi:2011st,Alonso:2012fy,Alonso:2013mca,Alonso:2013nca}, while the Abelian part can play a role in fixing the overall mass scales~\cite{Alonso:2011jd,Arias-Aragon:2017eww}. For the latter, there is not a unique possibility, but the requirement of minimality, in terms of scalar fields to be added to the spectrum, leads to the identification of $\UPQ$ as the best solution. We will therefore focus on such scenario for the rest of the paper. 

In order to guarantee the invariance under the flavour symmetry of the entire Lagrangian, Yukawa and mass matrices are promoted to spurion fields, transforming only under the flavour symmetry. The generic LSS Lagrangian for the lepton sector reads as follows:
\be
\begin{split}
-\sL_Y=&\,\ov{\ell_L}\,H\,\cY_e\, e_R+
\ov{\ell_L}\,\tH\, \cY_\nu\, N_R+
\epsilon\,\ov{\ell_L}\,\tH\, \cY'_\nu\, S_R\,+\\
&+\dfrac12\,\ov{N_R^c}\,\mu'\,N_R+
\dfrac12\,\ov{S_R^c}\,\mu\, S_R+
\dfrac12\left(\ov{N_R^c}\,\Lambda\,S_R+\ov{S_R^c}\,\Lambda^T\,N_R\right)
+\hc\,,
\end{split}
\label{GenericLagISS}
\ee
that matches Eq.~\eqref{genericISSmatrix}. These operators are invariant under $\cG_F^\text{NA}$ only if the six spurions transform as
\be
\begin{aligned}
\cY_e\sim({\bf3},\ov{\bf3},1,1)\,,\qquad
&&\cY_\nu\sim({\bf3},1,\ov{\bf3},1)\,,\qquad
&&\cY'_\nu\sim({\bf3},1,1,\ov{\bf3})\,,\\[2mm]
\mu'\sim(1,1,1,\ov{\bf6})\,,\qquad
&&\mu\sim(1,1,\ov{\bf6},1)\,,\qquad
&&\Lambda\sim(1,1,\ov{\bf3},\ov{\bf3})\,.
\end{aligned}
\ee
As can be deduced by looking at Eq.~\eqref{genericISSmnu}, not all these spurions are equally relevant for the active neutrino mass matrix; indeed, $\mu'$ does not contribute at tree level. Moreover, if one of the two terms in the $m_\nu$ expression dominates, then neutrino masses and lepton mixing can be determined in terms of only three spurions, $\Lambda$, $\cY_\nu$ and either $\mu$ or $\cY'_\nu$. However, this is not sufficient to achieve the predictive power desirable within the MFV context. To improve in this regard, the first simplification that can be adopted is to identify the $SU(3)_{N_R}$ and $SU(3)_{S_R}$ groups. In doing so, there is no need to introduce neither two separate neutrino Dirac Yukawa spurions $\cY_\nu$ and $\cY'_\nu$, nor two distinct Majorana mass terms $\mu$ and $\mu'$. Furthermore, the Majorana mixed mass term $\Lambda$ would have the same transformation properties of $\mu$ and $\mu'$. This leads to a simplified scenario where
\be
\begin{gathered}
SU(3)_{N_R}\times SU(3)_{S_R}\longrightarrow SU(3)_{N_R+S_R}\,,\\[2mm]
\cY_\nu\sim\cY'_\nu\sim({\bf3},1,\ov{\bf3})\,,\qquad
\mu\sim\mu'\sim\Lambda\sim(1,1,\ov{\bf6})\,,
\end{gathered}
\ee
that resembles the setup of the type-I Seesaw mechanism, with two neutrino spurions. As summarised in the introduction, even this setup is not predictive enough. To overcome this problem, we identified three possible frameworks.
\begin{description}
\item[CASE A:] The $SU(3)_{N_R+S_R}$ group is further reduced to its orthogonal version and the final flavour group reads
\be
\cG_F^\text{NA}=SU(3)_{\ell_L}\times SU(3)_{e_R}\times SO(3)_{N_R+S_R}\,.
\label{FsymCaseA}
\ee
The mass terms are proportional to the identity matrix, 
\be
\mu,\,\mu'\,,\Lambda\propto\unity\,,
\ee
while the only remaining neutrino spurion with non-trivial flavour structure transforms as
\be
\cY_\nu\sim\cY'_\nu\sim({\bf3},1,\ov{\bf3})\,.
\ee
\item[CASE B:] The symmetry groups associated to the EW lepton doublets and to the exotic neutral leptons are identified to a vectorial unitary group, such that the complete non-Abelian symmetry in the lepton sector reduces to
\be
\cG_F^\text{NA}=SU(3)_{V}\times SU(3)_{e_R}\,.
\ee
Both $N_R$ and $S_R$ transform as $\bf3$ under the vectorial group. In this case, Schur's Lemma guarantees that the neutrino Dirac Yukawa matrices are singlets of the symmetry group~\cite{Bertuzzo:2009im,AristizabalSierra:2009ex} and therefore $\cY_\nu$ and $\cY'_\nu$ are proportional to the identity matrix. The only remaining neutrino spurion with non-trivial flavour structure is the mass, that transforms as
\be
\mu\sim\mu'\sim\Lambda\sim(\ov{\bf6},1)\,.
\ee
\item[CASE C:] The flavour symmetry of this model is the same as the previous one, but in this case $N_R$ and $S_R$ transform differently: $N_R\sim({\bf3},1)$ and $S_R\sim(\ov{\bf3},1)$. Also in this case, the application of Schur's Lemma reduces the total number of spurions with non-trivial flavour structure, but with a different structure with respect to the previous scenario:
\be
\cY_\nu,\,\Lambda\propto\unity\,,\qquad\qquad
\mu'\sim\cY^{\prime\dagger}_\nu\sim\mu^\dag\sim(\ov{\bf6},1)\,.
\ee
\end{description}

Despite the differences in the flavour symmetries and in the field transformations, all three realisations describe correctly the lepton masses and mixing once the spurions acquire specific background values. The latter depend on the specific realisation, giving rise to different associated phenomenologies. We will discuss these details in Sects.~\ref{sec:CASEA}, \ref{sec:CASEB} and \ref{sec:CASEC}.

We will now discuss the implementation of the $\UPQ$ symmetry in order to explain the suppression of the tau (and bottom quark) mass with respect to the top mass and the smallness of the $\mu$, $\mu'$ and $\epsilon$ parameters in the neutral lepton mass matrix, as required for the LSS scenario. This discussion applies to the three cases identified above. 

Any fermion $\psi$ transforms under $\UPQ$ with a charge $x_\psi$. To ensure the invariance of the Lagrangian under this symmetry, we introduce a scalar field $\Phi$ that transforms only under $\UPQ$ and, without any loss of generality, fix its charge to be $x_\Phi=-1$. A generic Yukawa term can then be written as a non-renormalisable operator suppressed by powers of the cut-off scale $\Lambda_\Phi$:
\be
y_\psi\,\ov{\psi_L}\,H\,\psi_R\rightarrow y_\psi\,\ov{\psi_L}\,H\,\psi_R\left(\dfrac{\Phi}{\Lambda_{\phi}}\right)^{x_{\psi_R}-x_{\psi_L}}\,.
\label{GenericYukawaFN}
\ee
Similarly to what occurs in the Froggatt-Nielsen flavour model~\cite{Froggatt:1978nt}, once the scalar field develops a VEV, $\vev\Phi\equiv v_\Phi/\sqrt2$, and the EW symmetry breaking takes place, a fermion mass arises from this term. It is useful to introduce a parameter to indicate the ratio between this VEV and the cut-off scale:
\be
\vep\equiv\dfrac{v_\Phi}{\sqrt2\Lambda_\Phi}\,.
\ee
The mass term corresponding to the generic Yukawa term then reads
\be
m_\psi=y_\psi\,\dfrac{v}{\sqrt2}\,\vep^{x_{\psi_R}-x_{\psi_L}}\,.
\ee

As discussed above, in the M(L)FV context, the Abelian factors of the flavour symmetry only fix the different overall scales of the Yukawa matrices of each particle species. Indeed, fermions of the same species transform with the same charge under $\UPQ$: the three right-handed (RH) charged leptons transform with $x_e$, the three left-handed (LH) lepton doublets with $x_\ell$ and so on.

We can now proceed to fix as many values of the PQ charges as possible. As the top quark Yukawa coupling is close to $1$, then the whole Yukawa term for the up-type quarks should arise at the renormalisable level: the only PQ charge choice consistent with this requirement is
\be
x_u-x_Q=0\,.
\label{cau=0}
\ee
From the comparison of the bottom and tau masses, $m_b$ and $m_\tau$, with the top mass, $m_t$, two additional requirements follow:
\be
x_d-x_u\simeq\log_\vep(m_b/m_t)\,,\qquad\qquad
x_e-x_\ell\simeq\log_\vep(m_\tau/m_t)\,.
\label{xchargesdefinition}
\ee
The exact value of $\vep$ depends on the specific ultraviolet theory that gives rise to the effective Yukawa operators as in Eq.~\eqref{GenericYukawaFN}. Considering that $\vep<1$ and that $v_\Phi$ is expected, due to naturalness, to remain in the ballpark of $\Lambda_\Phi$, the interval of values that will be considered in this paper is $\vep\in[0.01,\,0.3]$, which is consistent with previous studies in the Froggatt-Nielsen approach. The presence of the logarithm in Eq.~\eqref{xchargesdefinition} softens the dependence on the exact value of $\vep$: to fix two benchmark values that will be used in the phenomenological analysis, we choose
\be
x_d-x_u=x_e-x_\ell=
\begin{cases}
1 \quad\text{for}\quad \vep=0.01\,,\\[2mm]
3 \quad\text{for}\quad \vep=0.23\,,
\end{cases}
\label{TwoCasesEpsilon}
\ee
where the first equality follows from the closeness of the bottom quark and tau masses, while the second benchmark with $\vep=0.23$ corresponds to the Cabibbo angle, as traditionally considered in the Froggatt-Nielsen framework.

The PQ charges for the exotic neutral leptons can be fixed by imposing the LSS structure of the neutral lepton mass matrix. The smallness of the entries proportional to $\mu^{(\prime)}$ in favour of the entries with $\Lambda$ may naturally follow from the condition $x_{N_R}=-x_{S_R}$. Moreover, in order to suppress the entries with $\cY'_\nu$ with respect to the entries with $\cY_\nu$, we can similarly impose $x_\ell=x_{N_R}$, thus identifying PQ with lepton number.
As we will show in the next sections, additional constraints on the values of the PQ charges follow from considering bounds on 
several observables, such as the radiative rare charged lepton decays, $\mu\to e$ conversion in nuclei, the effective number of neutrinos $N_\nu$ and the $W$ boson mass $M_W$.

We conclude this section discussing the consequences of the spontaneous breaking of the PQ symmetry. In the broken phase, the complex scalar field $\Phi$ can be written as
\be
\Phi=\dfrac{\rho+v_\Phi}{\sqrt2}e^{i\,a/v_\Phi}\,,
\ee
where $\rho$ is the radial component and $a$, which can be identified with the Majoraxion, is the angular one. As discussed in Ref.~\cite{Arias-Aragon:2017eww}, the VEV $v_\Phi$ has to be large and as a result the mass induced for the $\rho$ component is much larger than the EW scale (see  for details on the associated scalar potential). Therefore, it can be safely integrated out, leaving the Majoraxion as the only light degree of freedom relevant at the energies discussed here. The low-energy Lagrangian containing the Majoraxion interactions reads
\be
\sL_a=\dfrac12\derp_\mu a\derp^\mu a-c_{a\psi}\dfrac{\derp_\mu a}{2v_\Phi}\ov{\psi}\gamma^\mu\gamma_5\psi-\dfrac{\alpha_X}{8\pi}c_{aXX'}X^{(a)\mu\nu}X^{\prime(a)}_{\mu\nu}\,,
\ee
where $X^{(a)}_{\mu\nu}=\derp_\mu X^{(a)}_\nu-\derp_\nu X^{(a)}_\mu$  for each gauge boson $X$ in the mass basis, with the index $a$ present only for gluons, $\alpha_X=g_X^2/4\pi$ with $g_X$ the associated gauge coupling, and the $c_i$ being coefficients that can be linked to the PQ charges. In particular,
\be
c_{au}=x_Q-x_u\,,\qquad\qquad
c_{ad}=x_Q-x_d\,,\qquad\qquad
c_{ae}=x_\ell-x_e\,,
\ee
while the couplings with the gauge boson field strengths originate at the quantum level due to the non-vanishing of the derivative of the axial current. See Ref.~\cite{Arias-Aragon:2017eww} for the explicit expressions in the context we are considering.

It is useful to introduce the scale $f_a$, that is the one usually associated to the QCD axion in the literature: 
\be
f_a\equiv \dfrac{v_\Phi}{c_{agg}}\,,
\ee
that depends on the PQ charges of the quarks through the presence of $c_{agg}$ in its definition, where $c_{agg}=3(c_{au}+c_{ad})$.

Many different bounds are present on the scale $f_a$, from both astrophysical and terrestrial observables (see Refs.~\cite{Raffelt:2006cw,Mimasu:2014nea,Vinyoles:2015aba,DiLuzio:2016sbl,Brivio:2017ije,Dolan:2017osp,Bauer:2017ris,Arias-Aragon:2017eww,Alonso-Alvarez:2018irt,Bauer:2018uxu,Gavela:2019wzg,Merlo:2019anv,Arias-Aragon:2020qtn,Arias-Aragon:2020shv} for a very comprehensive, but still incomplete list). The strongest one in this context is the constraint from the observation of Red Giants cooling~\cite{Viaux:2013lha}: for masses of the Majoraxion $m_a\lesssim1\eV$,
\be
\dfrac{c_{ae}}{c_{agg}f_a}=\dfrac{1}{3f_a}\lesssim 8.6\times 10^{-10}\GeV^{-1}\,.
\ee
The first equality follows from Eq.~\eqref{cau=0} and from the closeness of the bottom quark and the tau lepton masses.
The corresponding bound on the Majoraxion scale $f_a$ reads
\be
f_a\gtrsim 3.9\times 10^8\GeV\qquad\longleftrightarrow\qquad
\begin{cases}
\vep=0.01 & \longrightarrow v_\Phi\gtrsim1.2\times 10^{9}\GeV\,,\\[2mm]
\vep=0.23 & \longrightarrow v_\Phi\gtrsim3.5\times 10^{9}\GeV\,.
\end{cases}
\ee

It is interesting to notice that for values of the scale $f_a$ around $10^9\GeV$, the Majoraxion may be a valid DM candidate. Indeed, while the well-known misalignment mechanism~\cite{Preskill:1982cy,Abbott:1982af,Dine:1982ah} requires a larger $f_a$ for generic QCD axions to be produced in the right amount to be DM, topological defects may provide an additional source of cold axions: when the PQ symmetry is broken after inflation, domain walls and cosmic strings associated to its breaking form and may remain in our Universe. Their decay leads to the production of cold axions and, despite the theoretical uncertainties existing associated to this production mechanism, the analyses show that $f_a \gtrsim 10^9 \GeV$ reproduces the correct relic density of DM axions~\cite{Gorghetto:2018myk,Gorghetto:2020qws}.

\boldmath
\section{$\cG_F^\text{NA}=SU(3)_{\ell_L}\times SU(3)_{e_R}\times SO(3)_{N_R+S_R}$}
\unboldmath
\label{sec:CASEA}

This section is devoted to the first scenario we outlined in the previous section. The complete flavour symmetry characterising this framework is 
\be
\cG_F^{SO(3)}\equiv SU(3)_{\ell_L}\times SU(3)_{e_R}\times SO(3)_{N_R+S_R}\times \UPQ\,,
\ee
and the transformation properties under it of the involved fields can be read in Tab.~\ref{Tab.CaseA}. Interestingly, in this scenario, the number of exotic neutral leptons can be different from 3. In Ref.~\cite{Gavela:2009cd}, it is shown that the minimal scenario compatible with data from neutrino oscillations contains only one $N_R$ and one $S_R$. Although in the rest of this section we will only refer to the $3+3$ scenario, many of the results can be easily extended to the $n+n$ case, with $n>1$.

\begin{table}[h!]
\centering
\begin{tabular}{c|ccc|c} 
&  $SU(3)_{\ell_L} $ & $SU(3)_{e_R}$ & $SO(3)_{N_R+S_R}$ & $U(1)_{PQ}$\\ 
\hline 
\hline 
&&&&\\[-3mm]
$\ell_L$  	& $\bf 3$ 		& 1		    & 1		    & $x_\ell$\\
$e_R$   	& 1  			& $\bf 3$	& 1		    & $x_e$ \\
$N_R$  		& 1  		    & 1 		& $\bf 3$   & $x_\ell$\\
$S_R$    	& 1  			& 1 		& $\bf 3$	& $-x_\ell$\\
$\Phi$  	& 1 			& 1 		& 1 		& $-1$ \\
\hline
&&&&\\[-3mm]
$Y_e$   	& $\bf3$  		& $\bf\ov3$	& 1		    & 0\\
$Y_\nu$	    & $\bf3$ 		& 1  		& $\bf3$	& 0\\
\end{tabular}
\caption{\em Transformation properties of the SM leptons, exotic neutral leptons and spurions under the global symmetries $\cG_F^{SO(3)}$.}
\label{Tab.CaseA}
\end{table}

The mass Lagrangian invariant under this symmetry can be written as 
\begin{align}
-\sL^\text{A}_Y=&\phantom{+}\ov{\ell_L}HY_ee_R\left(\dfrac{\Phi}{\Lambda_\Phi}\right)^{x_e-x_\ell}+
\ov{\ell_L}\tH Y_\nu N_R+
c_\nu\ov{\ell_L}\tH Y_\nu S_R\left(\dfrac{\Phi}{\Lambda_\Phi}\right)^{2x_\ell}+\\
&+\dfrac12c_N\ov{N_R^c}N_R\Phi\left(\dfrac{\Phi}{\Lambda_\Phi}\right)^{2x_\ell-1}+
\dfrac12c_S\ov{S_R^c}S_R\Phi^\dag\left(\dfrac{\Phi^\dag}{\Lambda_\Phi}\right)^{2x_\ell-1}+
\Lambda\ov{N_R^c}S_R+\hc\,,\nn
\end{align}
where $c_\nu$, $c_N$, and $c_S$ are free real parameters and $\Lambda$ is a real scale. In particular, the associated terms do not present any non-trivial flavour structure. The only flavour information is contained in $Y_\nu$, that appears in both Dirac Yukawa terms. Notice the change in the notation with respect to the Lagrangian in Eq.~\eqref{GenericLagISS}: the generic $\cY_\nu$ and $\cY'_\nu$ are now substituted by the spurion $Y_\nu$, and powers of $\Phi/\Lambda_\Phi$ play the role of the generic suppressing parameters $\epsilon$ and $\mu^{(\prime)}$.

After flavour and EW symmetry breaking, the associated neutral mass matrix in the notation of Eq.~\eqref{GenericNeutralMassLag} reads
\be
\cM_\chi=
\left(
\begin{array}{ccc}
0  
& \dfrac{v}{\sqrt2}Y_\nu  
& c_\nu\dfrac{v}{\sqrt2}\vep^{2x_\ell}Y_\nu  \\[4mm]
\dfrac{v}{\sqrt2}Y^T_\nu  
& c_N\dfrac{v_\Phi}{\sqrt2}\vep^{2x_\ell-1}  
& \Lambda  \\[4mm]
c_\nu\dfrac{v}{\sqrt2}\vep^{2x_\ell}Y^T_\nu  
& \Lambda 
& c_S\dfrac{v_\Phi}{\sqrt2}\vep^{2x_\ell-1}  
\end{array}
\right)\,,
\ee
where the blocks $22$ and $33$ are suppressed with respect to the block $23$ due to the presence of powers of $\vep$, and the same occurs to the block $13$ in favour of the block $12$. This enforces the LSS structure discussed in the introduction. By block diagonalising the mass matrix, we can identify the mass eigenvalues: at leading order,
\be
m_\nu=\dfrac{v^2}{2}\vep^{2x_\ell-1}\left(\dfrac{c_S v_\Phi}{\sqrt2\Lambda^2}-\dfrac{2c_\nu\vep}{\Lambda}\right)Y_\nu Y_\nu^T\,,\qquad\qquad
M_N\simeq \Lambda\,,
\label{CaseANeuMassMatrices}
\ee
that refer respectively to the light active neutrinos and to the heavy neutral leptons, which are degenerate at this level of approximation. The new intermediate states, where the mass matrix $\cM_\chi$ is block diagonal, are given by $\nu_L'\simeq\nu_L-\Theta S_R^c$, where
\be
\Theta\simeq\dfrac{v}{\sqrt2\Lambda}Y_\nu\,.
\label{ThetaCaseA}
\ee

The next step consists in writing the spurions in terms of lepton masses and the entries of the PMNS matrix. Without any loss of generality, it is possible to perform flavour transformations such that the charged lepton mass matrix is diagonal in the flavour basis and, as a result, in the charged lepton sector
\be
Y_e=\dfrac{\sqrt2}{v}\widehat{M}_e\equiv\dfrac{\sqrt2}{v}\diag(m_e,\,m_\mu,\,m_\tau)\,.
\label{ChargedLeptonSpurion}
\ee
In this basis, the active neutrino mass matrix can be diagonalised by a unitary matrix, $U$, defined as
\be
\widehat{m}_\nu=U^\dag m_\nu U^\ast\,,
\label{DiagonalisationOfmnu}
\ee
where $\widehat{m}_\nu$ is the diagonal active neutrino mass matrix, $\widehat{m}_\nu\equiv\diag\left(m_1,\,m_2,\,m_3\right)$. Notice that the matrix that appears in charged current interactions (which we will call the PMNS mixing matrix), will be the product of the unitary $U$ times a non-unitary matrix from the previous block-diagonalisation through $\Theta$. This correction is however bounded to be small~\cite{Antusch:2014woa,Fernandez-Martinez:2016lgt} and can be neglected for neutrino oscillation phenomenology, such that the $U$ matrix is given by
\be
U\equiv R_{23}(\theta_{23})\cdot R_{13}(\theta_{13},\,\delta_\text{CP})\cdot R_{12}(\theta_{12})\cdot\diag\left(1,\,e^{i\,\alpha_{2}},\,e^{i\,\alpha_{3}}\right)\,,
\label{DefinitionPMNS}
\ee 
with $R_{ij}(\theta_{ij})$ a generic rotation of angle $\theta_{ij}\in[0,\,\pi/2]$ in the $ij$ sector, with the addition of the Dirac CP phase $\delta_\text{CP}\in[-\pi,\,\pi]$ in the reactor sector. $\alpha_{2,3}\in[0,\pi]$ are the Majorana phases~\cite{deGouvea:2008nm}.

By inverting Eq.~\eqref{CaseANeuMassMatrices}, we obtain a constraint on the neutrino Yukawa spurion, such that
\be
Y_\nu\,Y_\nu^T=\dfrac{1}{f}\,U\,\widehat{m}_\nu\,U^T\,,
\ee
where
\be
f\equiv\dfrac{v^2\,\vep^{2x_\ell-1}}{2\,\sqrt2\,\Lambda^2}\left(c_S\,v_\Phi-2\,\sqrt2\,c_\nu\,\vep\,\Lambda\right)\,.
\label{fCASEA}
\ee
This condition translates into the following expression for $Y_\nu$:
\be
Y_\nu=\dfrac{1}{f^{1/2}} U \widehat{m}_\nu^{1/2} \cH^T\,,
\ee
where $\cH$~\footnote{In the Casas-Ibarra parameterisation~\cite{Casas:2001sr}, a generic complex orthogonal matrix, dubbed $R$, appears instead of $\cH$. However, decomposing $R$ as a product of $\cH$ and a real and orthogonal matrix, the latter can be absorbed by a transformation of the exotic neutral leptons, which are degenerate in mass~\cite{Cirigliano:2006nu}.} is a complex orthogonal and Hermitian matrix, $\cH^T\cH=\unity$ and $\cH^\dag=\cH$, that can be parameterised as
\be
\cH\equiv e^{i\phi}=\unity-\dfrac{\cosh r-1}{r^2}\phi^2+i\dfrac{\sinh r}{r}\phi\,,
\ee
with $\phi$ a matrix that depends on three additional real parameters,
\be
\phi=\left(
\begin{array}{ccc}
0
& \phi_1
& \phi_2\\
-\phi_1
& 0
& \phi_3\\
-\phi_2
& -\phi_3
& 0
\end{array}
\right)\,,
\ee
and $r\equiv\sqrt{\phi_1^2+\phi_2^2+\phi_3^2}$.
The presence of the $\cH$ matrix is a characteristic feature of this framework with respect to the following two that will be discussed in the next sections and leads to a more cumbersome phenomenological discussion. Indeed, the only $d=6$ operator generated at tree level by integrating out the exotic neutral leptons is the one introduced in Eq.~\eqref{Genericd6Operator}, with the dimensional Wilson coefficient given by
\be
c_{\cO_{d=6}}=\dfrac{Y_\nu Y_\nu^\dag}{\Lambda^2}=\frac{1}{f\Lambda^2}
U\,\widehat{m}_\nu^{1/2}\,\cH^T\,\cH^*\,\widehat{m}_\nu^{1/2}\,U^\dagger \,.
\ee
 The dependence on $\cH$ is therefore non-trivial and prevents the prediction of low-energy observables in terms of the neutral lepton masses, PMNS matrix elements and NP scales only. In the following, we will analyse the parameter space of these $\phi_i$ and their impact in several low-energy observables.

\subsection*{Phenomenological Consequences}

In this section we will focus on a set of precision electroweak and flavour observables that are affected by the presence of the exotic neutral leptons. In particular, we will discuss the radiative rare decay of the muon, muon conversion in nuclei, the effective number of neutrinos $N_\nu$ as determined by the invisible width of the $Z$ and the comparison of the measured $W$ boson mass $M_W$ with its prediction in terms of the Fermi constant from muon decay.

For convenience, we introduce the parameter $\eta$, that encodes the new contributions to the low-energy observables~\cite{Fernandez-Martinez:2007iaa}:
\be
\eta\equiv\dfrac12\Theta\Theta^\dag=\frac{v^2}{4\Lambda^2}Y_\nu Y_\nu^\dagger=\frac{v^2}{4f\Lambda^2}U\,\widehat{m}_\nu^{1/2}\,\cH^T\,\cH^*\,\widehat{m}_\nu^{1/2}\,U^\dag\,.
\ee
Notice that, in all generality, $\eta$ can parameterise any deviation from unitarity of the PMNS matrix $\cN$, that relates the active neutrinos to the mass basis:
\be
\cN=(\unity-\eta)\,U\,.
\ee

The analytical expressions for the observables listed above in terms of $\eta$ are well known in the literature (see Refs.~\cite{Langacker:1988ur,Antusch:2006vwa,Antusch:2014woa,Fernandez-Martinez:2016lgt}). In our analysis, the input parameters are the fine structure constant, the Fermi constant extracted from the dominant muon decay, and the $Z$ gauge boson mass. The corresponding values are~\cite{ParticleDataGroup:2020ssz}
\ba
\alpha_\text{em}=&7.2973525693(11)\times10^{-3},\\
G_\mu=&1.1663787(6)\times10^{-5}\GeV^{-2},\\
M_Z=&91.1876(21)\GeV\,.
\ea

The non-unitarity of the mixing matrix $\cN$ implies a modification of the dominant muon decay, whose decay rate reads\footnote{For simplicity, we write all SM expressions for the different observables at tree level. However, notice that SM loop corrections are numerically very relevant for these electroweak precision observables, and have been taken into account in the numerical analysis. Conversely, loop corrections involving the new heavy neutral leptons can be safely neglected~\cite{Fernandez-Martinez:2015hxa}.}
\be
\Gamma_\mu\simeq\dfrac{m_\mu^5G_F^2}{192\pi^3}\left(1-2\eta_{ee}-2\eta_{\mu\mu}\right)\equiv\dfrac{m_\mu^5G_\mu^2}{192\pi^3}\,,
\ee
implying the following relation between the Fermi constant parameter $G_F$, which enters the Fermi Lagrangian, and its experimental determination $G_\mu$:
\be
G_F=G_\mu\left(1+\eta_{ee}+\eta_{\mu\mu}\right)\,.
\ee

The relation between the $W$ boson mass and the experimental determination of $G_\mu$ through the muon decay, once including the non-unitarity corrections, reads
\be
    M_W = M_Z \sqrt{\frac{1}{2}+\sqrt{\frac{1}{4}-\frac{\pi\alpha_\mathrm{em}(1-\eta_{\mu\mu}-\eta_{ee})}{\sqrt{2}G_\mu M_Z^2}}}
\ee
while its experimental determination is $M_W=80.379(12)\GeV$~\cite{ParticleDataGroup:2020ssz} (we will comment on the new measurement at the CDF II detector in Sect.~\ref{sec:MWnew}). 

Another consequence of the non-unitarity of $\cN$ is the modification of the $Z$ gauge boson decay into neutrinos:
\be
\Gamma_\text{Z-inv}=\frac{G_\mu M_Z^3}{12\sqrt{2}\pi}(3-4\eta_{\tau\tau}-\eta_{ee}-\eta_{\mu\mu})\equiv \dfrac{G_\mu M_Z^3N_\nu}{12\sqrt{2}\pi}\,,
\ee
where $N_\nu$ is the number of active neutrinos. By comparing with the experimental determination, $N_\nu=2.9963(74)$~\cite{Janot:2019oyi}, it is possible to further constrain the $\eta_{\alpha \alpha}$ terms.

The last observables that we consider are the radiative rare charged lepton decay branching ratios and the ratio between the $\mu\to e$ conversion rate over the capture rate $\Gamma_\text{capt}$ in light nuclei. The branching ratio of the radiative processes, in the limit in which the heavy neutral lepton masses $M_N$ are much larger than the EW scale, generically reads
\be
\BR(\ell_i\to \ell_j\gamma)\equiv\dfrac{\Gamma(\ell_i\to \ell_j\gamma)}{\Gamma(\ell_i\to \ell_j\nu\ov\nu)}=\frac{3\alpha_\text{em}}{2\pi}\vert\eta_{\ell_j\ell_i}\vert^2\,.
\label{BRmutoegamma}
\ee
The analytic expression for the $\mu \to e$ conversion in nuclei reads~\cite{Alonso:2012ji}
\be
\begin{split}
R_{\mu\to e}=&{\scriptsize\frac{\sigma\left(\mu^- X\rightarrow e^- X\right)}{\sigma\left(\mu^- X\rightarrow \text{Capture}\right)}}\normalsize\\
\simeq&\frac{G_\mu^2 \alpha^5_\text{em}m_\mu^5}{2s_w^4\pi^4\Gamma_{\text{capt}}}
\frac{Z^4_{\text{eff}}}{Z}\vert\eta_{e\mu}\vert^2F_p^2\Big[\left(A+Z\right)F_u+\left(2A-Z\right)F_d \Big]^2\,,
\end{split}
\label{MutoeConversion}
\ee
where $X$ stands for the nucleus ${}^A_Z N$, $A$ stands for the atomic mass number, $Z$ ($Z_\text{eff}$) for the (effective) atomic number, and $F_p$ is a nuclear form factor~\cite{Kitano:2002mt,Suzuki:1987jf}, whose values are summarised in Tab.~\ref{Tab.MutoEConversionParameters} for each element. On the other hand, $F_u$ and $F_d$ are form factors associated to the neutrino physics parameters and are defined in~\cite{Alonso:2012ji,Fernandez-Martinez:2016lgt}:
\ba
F_u=&\dfrac23s^2_W\dfrac{16\log\left(\frac{M_N^2}{M_W^2}\right)-31}{12}-\dfrac{3+3\log\left(\frac{M_N^2}{M_W^2}\right)}{8}\,,\\
F_d=&-\dfrac13s^2_W\dfrac{16\log\left(\frac{M_N^2}{M_W^2}\right)-31}{12}-\dfrac{3-3\log\left(\frac{M_N^2}{M_W^2}\right)}{8}\,,
\ea
where $s^2_W$ can be taken at its SM predicted value, $s^2_W=0.22377(10)$ (on-shell scheme)~\cite{ParticleDataGroup:2020ssz}, as the non-unitarity  deviations would correspond to higher order corrections.

\begin{table}[t!]
\centering
\begin{tabular}{c|ccc} 
Nucleus ${}^A_Z N$ &  $Z_\text{eff} $ & $F_p$ & $\Gamma_\text{capt} (10^6\,\text{s}^{-1})$\\[1mm]
\hline 
\hline 
&&&\\[-4mm]
${}^{27}_{13}$ Al& $11.5$ & $0.64$ & $0.7054$\\[1mm]
${}^{48}_{22}$ Ti& $17.6$ & $0.54$ & $2.59$\\[1mm]
${}^{197}_{79}$ Au& $33.5$ & $0.16$ & $13.07$
\end{tabular}
\caption{\em Atomic mass number $A$, (effective) atomic number $Z$ ($Z_\text{eff}$), form factor $F_p$ and capture rate $\Gamma_\text{capt}$ for the three nuclei Al, Ti, Au. Values from Refs.~\cite{Suzuki:1987jf,Kitano:2002mt}.}
\label{Tab.MutoEConversionParameters}
\end{table}

We can now proceed with the numerical analysis in order to study the parameter space in terms of $x_\ell$ and the mass of the heavy neutral leptons. The numerical inputs, besides those already reported above, are the neutrino oscillation parameters in Tab.~\ref{Tab.NuOscillationParameters}, for both the normal ordering (NO) and inverted ordering (IO) for the active neutrino spectrum, taken from Ref.~\cite{Esteban:2020cvm} and fixed at their best fit central values. 

\begin{table}[h!]
\centering
\begin{tabular}{c|cc} 
Observable  &  Normal Ordering & Inverted Ordering \\[1mm]
\hline 
\hline 
&&\\[-4mm]
$\sin^2\theta_{12}$ & $0.304^{+0.012}_{-0.012}$ & $0.304^{+0.013}_{-0.012}$ \\[2mm]
$\sin^2\theta_{23}$ & $0.450^{+0.019}_{-0.016}$ & $0.570^{+0.016}_{-0.022}$ \\[2mm]
$\sin^2\theta_{13}$ & $0.02246^{+0.00062}_{-0.00062}$ & $0.02241^{+0.00074}_{-0.00062}$ \\[2mm]
$\delta_\text{CP}/{}^\circ$ & $-130^{+36}_{-25}$ & $-82^{+22}_{-30}$ \\[2mm]
$\dfrac{\Delta m_\text{sol}^2}{10^{-5}\eV^2}$ & $7.42^{+0.21}_{-0.20}$ & $7.42^{+0.21}_{-0.20}$ \\[2mm]
$\dfrac{\vert\Delta m_\text{atm}^2\vert}{10^{-3}\eV^2}$ & $2.510^{+0.027}_{-0.027}$ & $2.490^{+0.026}_{-0.028}$
\end{tabular}
\caption{\em Three-flavour oscillation parameters from Ref.~\cite{Esteban:2020cvm}. The mixing angles and the Dirac CP phase are defined in Eq.~\eqref{DefinitionPMNS}, while the mass squared differences are defined as $\Delta m_\text{sol}^2\equiv m_2^2-m_1^2$, $\Delta m_\text{atm}^2\equiv m_3^2-m_1^2>0$ for NO and $\Delta m_\text{atm}^2\equiv m_3^2-m_2^2<0$ for IO.}
\label{Tab.NuOscillationParameters}
\end{table}

While the lightest active neutrino mass, $m_1$ in NO and $m_3$ in IO, can vanish, the largest value it can take is limited by the present bound from the Planck collaboration~\cite{Planck:2018vyg}: at $95\%$ C.L.,
\be
\sum m_i<0.12\eV\,,\qquad\qquad\text{Planck TT, TE, EE + lowE + lensing + BAO}
\label{SumNuMasses}
\ee
obtained considering the 2018 Planck data for the temperature power spectra (TT), the high-multipole TE and EE polarization spectra, the polarization data at low multipoles (lowE), the cosmic microwave background lensing (lensing) and the baryon acoustic oscillation measurements (BAO).
As a result, we will consider in our analysis that 
\be
m_1\leq30\meV\,,\qquad\qquad
m_3\leq16\meV\,,
\label{LargestValuesm1m3}
\ee
for NO and IO, respectively.
To conclude, the experimental bounds on the lepton flavour violating observables used in the analysis can be found in Tab.~\ref{Tab.ExperimentalBounds}.

\begin{table}[h!]
\centering
\begin{tabular}{c|cc} 
Observable  & Experimental Bound & Future Sensitivity\\[1mm]
\hline 
\hline 
&&\\[-4mm]
$\BR(\mu\to e \gamma)$ &$4.2\times10^{-13}$~\cite{ParticleDataGroup:2020ssz}& $6\times 10^{-14}$~\cite{Baldini:2021kfb} \\[2mm]
$\BR(\tau\to e \gamma)$ & $1.9\times 10^{-7}$ & $9\times10^{-9}$~\cite{SnowmassCLFVTau}\\[2mm]
$\BR(\tau\to \mu \gamma)$ & $2.5\times 10^{-7}$ & $6.9\times10^{-9}$~\cite{SnowmassCLFVTau}\\[2mm]
$R_{\mu\to e}\,(\text{Al})$& -- &$6\times10^{-17}$~\cite{Kutschke:2011ux}\\[2mm]
$R_{\mu\to e}\,(\text{Ti})$&$4.3\times10^{-12}$~\cite{ParticleDataGroup:2020ssz} &$10^{-18}$ ~\cite{Barlow:2011zza}\\[2mm]
$R_{\mu\to e}\,(\text{Au})$&$7\times10^{-13}$~\cite{ParticleDataGroup:2020ssz}& --
\end{tabular}
\caption{\em Experimental determinations and future sensitivities, at $90\%$C.L., for a selected list of lepton flavour violating processes. ``--'' indicates that no bound is expected. The values of $\BR(\tau\to e \gamma)$ and $\BR(\tau\to \mu \gamma)$ have been obtained according to the definition in Eq.~\eqref{BRmutoegamma}, with experimental data from Ref.~\cite{ParticleDataGroup:2020ssz}.}
\label{Tab.ExperimentalBounds}
\end{table}

An additional constraint that strongly reduces the parameter space is associated to the consistency of the theory: we impose that the spurions always remain in the perturbative regime. While this is trivially satisfied by $Y_e$, as the largest entry is the one associated to the tau and is $\sim0.01$, the neutrino Dirac Yukawa spurion may take extremely large values due to the matrix $\cH$. We therefore exclude the parameter space where any of its entries is larger than 1 in modulus, that is $\left|\max\left(Y_\nu\right)\right|>1$.

To simplify the analysis, we will take all the free coefficients entering $\eta$, and expected to be of order 1, exactly equal to $c_\nu=c_S=1$. Notice that the parameter $c_N$ does not affect the analysis at the level of approximation considered. 

\begin{figure}[h!]
    \centering
    \includegraphics[width=0.406\textwidth,trim=1 1 67 1, clip]{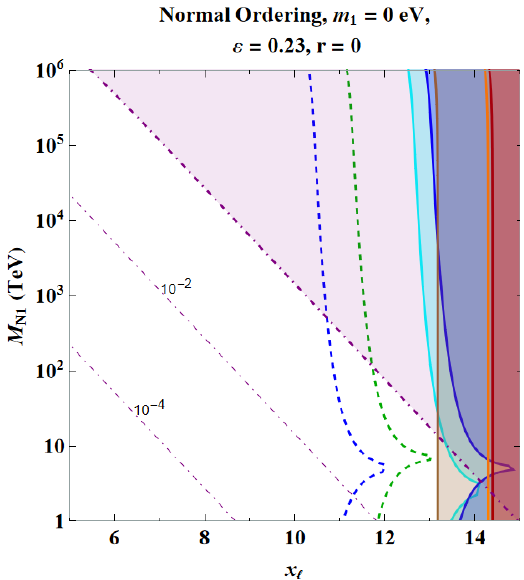}
    \includegraphics[width=0.584\textwidth,trim=1 1.3 1 1, clip]{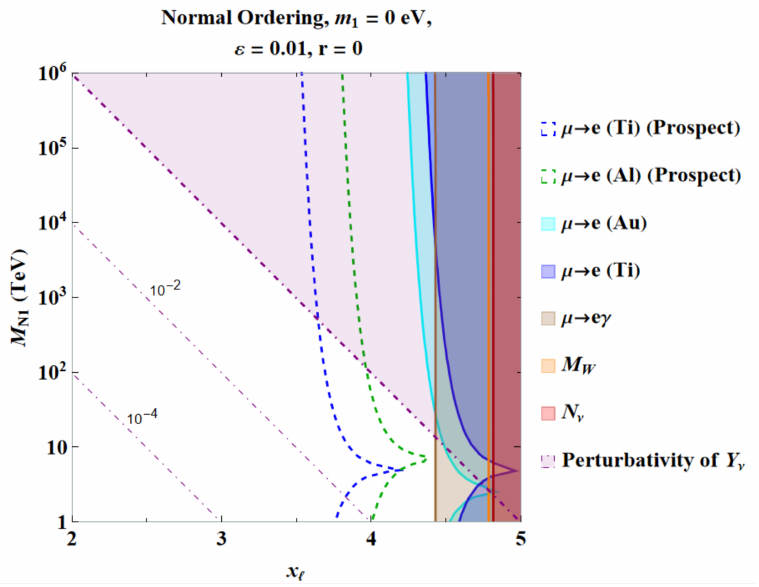}
    \caption{\em Parameter space of the mass of the lightest exotic neutral lepton $M_{N1}$ as a function of the PQ charge $x_\ell$. Lines and colours are explained in the legend and in the text. Both plots refer to the NO case with a massless lightest active neutrino $m_1=0$, assuming $\phi_i=0$, i.e. $r\equiv\sqrt{\phi_1^2+\phi_2^2+\phi_3^2}=0$. The plot in the left (right) holds for $\vep=0.23$ ($\vep=0.01$).}
    \label{fig:CaseA_MFV}
\end{figure}

Fig.~\ref{fig:CaseA_MFV} provides some intuition of the parameter space of this scenario in the specific case with $\phi_i=0$.  The plots show the allowed regions of the PQ charge $x_\ell$ and the mass of the lightest exotic neutral lepton $M_{N1}$. They refer to NO for the active neutrino mass spectrum, with the lightest neutrino $\nu_1$ with vanishing mass, and fixing the PQ expanding parameter to $\vep=0.23$ ($\vep=0.01$) on the left (right). Notice that letting the lightest active neutrino mass be different from zero or considering IO does not modify considerably the plots in Fig.~\ref{fig:CaseA_MFV}. Moreover, as $\cH=\unity$, the dependence on the Majorana phases disappears.

The shaded areas are excluded due to the experimental present data at $90\%$ C.L. and to the perturbativity of $Y_\nu$: in red the number of active neutrinos $N_\nu$, in orange the $W$ boson mass, in brown the $\BR(\mu\to e \gamma)$, in blue and cyan the $\mu\to e$ conversion in titanium and gold, respectively; the perturbativity bound on $Y_\nu$ is in light purple. The thick dashed lines in green and blue represent the future prospects for $\mu\to e$ conversion in aluminium and titanium, respectively. We do not show the future prospect for $\mu\to e\gamma$, as it is not competitive with the ones just mentioned. The thin dot-dashed purple lines indicate when the absolute value of the largest entry of $Y_\nu$ is equal to $10^{-2}$ and $10^{-4}$. One interesting feature of these plots is the cusp in the experimental bounds on $\mu\to e$ conversion in nuclei: the precise value of $M_{N1}$ at which the cusp appears depends on the nuclei, but in all the cases is in the interval $[5,\,10]\TeV$. This has been explained in Ref.~\cite{Alonso:2012ji} and is due to the cancellation that occurs in the parenthesis in Eq.~\eqref{MutoeConversion} between $F_u$ and $F_d$, that depend on $M_{N1}$.

Comparing the two plots, changing the value of $\vep$ from $0.23$ to $0.01$ has as major impact only a rescaling of $x_\ell$. This is due to the fact that the dominant term, see for example the left piece within the brackets in Eq.~\eqref{fCASEA}, depends on the combination $\vep^{2 x_\ell -1}$, and therefore for a larger $\vep$ corresponds a larger $x_\ell$ to keep the exponential constant. The information that can be extracted from these plots is that $x_\ell\lesssim13$ ($x_\ell\lesssim 4$) for $\vep=0.23$ ($\vep=0.01$) due to the present bounds on $\BR(\mu\to e\gamma)$. On the other side, the perturbativity of $Y_\nu$ puts upper bounds on the largest value of the lightest exotic neutral lepton. Additionally, if one requires that the largest absolute value of the entries of $Y_\nu$ is not too small, say not smaller than $10^{-2}$, the viable parameter space is the one between the thick and the first thin purple dot-dashed lines. According to this criterion, taking $\vep=0.23$, for the largest allowed value of the PQ charge $x_\ell=13$, the lightest exotic neutral lepton can have a mass in the range $M_{N1}\in[0.2,17]\TeV$ approximately, being the lowest values slightly above the present sensitivity at colliders (see Ref.~\cite{Abdullahi:2022jlv} and references therein). Lowering the value of $x_\ell$ translates into an increase of $M_{N1}$, making harder the possibility of their direct detection at colliders in the future. Very similar conclusions hold for $\vep=0.01$, once rescaling $x_\ell$. 

\begin{figure}[h!]
    \centering
    \includegraphics[width=0.4025\textwidth,trim=1 1 54 1, clip]{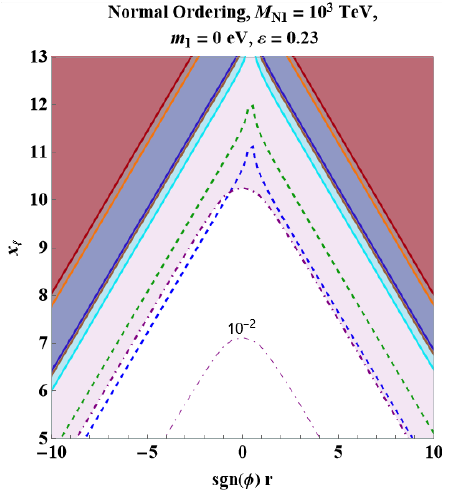}
    \includegraphics[width=0.575\textwidth,trim=1 2 1 1, clip]{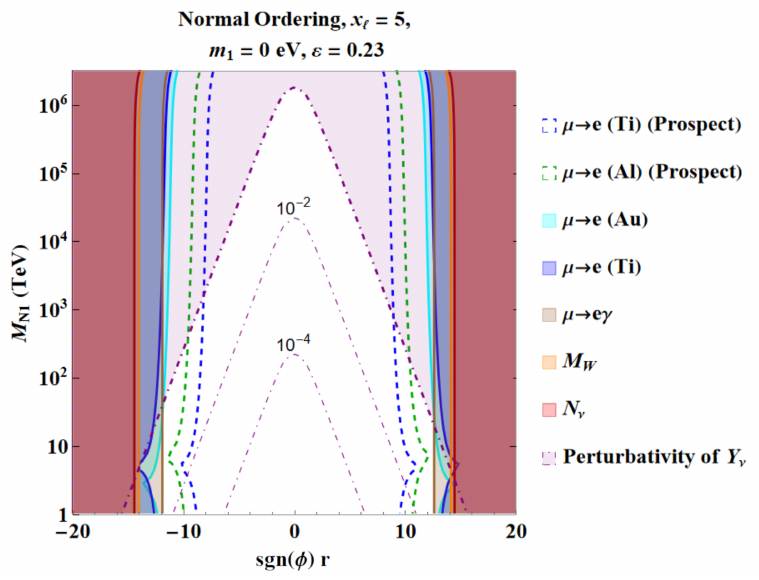}
    \caption{\em Parameter space of $x_\ell$ having fixed $M_{N1}=10^3\TeV$ (on the left) and of $M_{N1}$ having fixed $x_\ell=5$ (on the right) as a function of the parameter $\sign{\phi}r$. Both plots refer to the NO case with $m_1=0$, $\alpha_2=\alpha_3=0$, and assuming $\phi_1=\phi_2=\phi_3\equiv\phi$. Lines and colours are explained in the legend.}
    \label{fig:CaseA_Benchmarks}
\end{figure}

In Fig.~\ref{fig:CaseA_Benchmarks} we generalise the previous analysis, allowing for non-vanishing values of $\phi_i$, but taking them all equal, $\phi_1=\phi_2=\phi_3\equiv\phi$. In this case, there would be dependence on the Majorana phases, but their impact is negligible and are fixed to be vanishing for simplicity. In this way, we can discuss only the impact of the $\cH$ matrix, and enlarging the parameter space by only one dimension. In the plots in Fig.~\ref{fig:CaseA_Benchmarks} we show the parameter space for $x_\ell$ having fixed $M_{N1}=10^3\TeV$ (on the left) and for $M_{N1}$ having fixed $x_\ell=5$ (on the right) with respect to $r$. We also are taking into consideration positive and negative values of $\phi$, represented in the horizontal axis by the $\sign{\phi}$. Only the NO case with $m_1=0$ and $\vep=0.23$ is considered, as a non-vanishing lightest active neutrino mass does not have any significant effect and the IO scenario is very similar. Moreover, selecting $\vep=0.01$ amounts to a rescaling of the values of $x_\ell$ as seen in Fig.~\ref{fig:CaseA_MFV}.

The plot on the left of Fig.~\ref{fig:CaseA_Benchmarks} shows clearly that increasing the value of $r$ translates into a larger neutrino Yukawa spurion. As a result the perturbativity bound represents a very strong constraint, actually stronger than the present experimental bounds. For this value of $M_{N1}$, $\left|r\right|$ has to be smaller than $\sim8$. Interestingly,  when imposing a ``natural'' value of $Y_\nu$, that is $\left|\max\left(Y_\nu\right)\right|\in[10^{-2},\,1]$, small values of $x_\ell$ require non-vanishing values of $r$. Nevertheless, values of $r>1$ imply, by construction, that $Y_\nu Y_\nu^\dagger$ is significantly larger than $Y_\nu Y_\nu^T$. In other words, that the contribution to the $d=6$ operator is further enhanced with respect to that of neutrino masses, besides the $\vep^{2 x_\ell -1}$ suppression arising from the LSS. A very particular texture for $Y_\nu$, maybe originated by some additional symmetry, would be implied in this part of the parameter space.

The plot on the right of Fig.~\ref{fig:CaseA_Benchmarks} is very similar, although with the difference that the perturbativity condition dominates at smaller $\left|r\right|$, while the present experimental bounds do for larger $\left|r\right|$. Having fixed $x_\ell=5$, the allowed range of values is $\left|r\right|<13$. For larger values of the exotic neutral lepton mass, the allowed parameter space for $\left|r\right|$ gets shrunk. Also in this plot, we can appreciate that non-vanishing values of $r$ can be necessary in order to remain in the ``natural'' region for $Y_\nu$.

\begin{figure}[h!]
    \centering
    \includegraphics[width=0.48\textwidth,trim=2 2 2 2, clip]{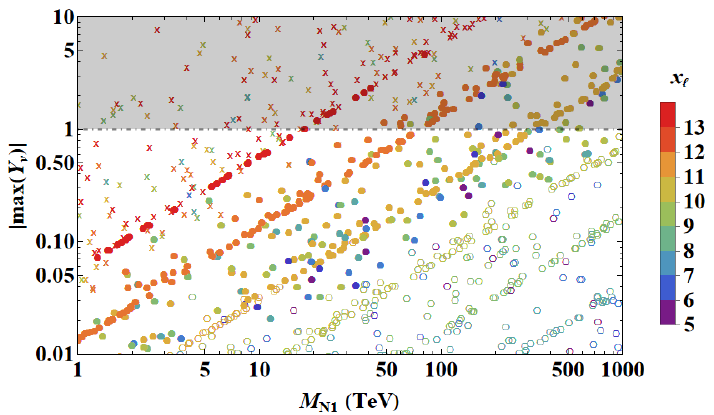}
    \includegraphics[width=0.49\textwidth,trim=2 2 1.25 2, clip]{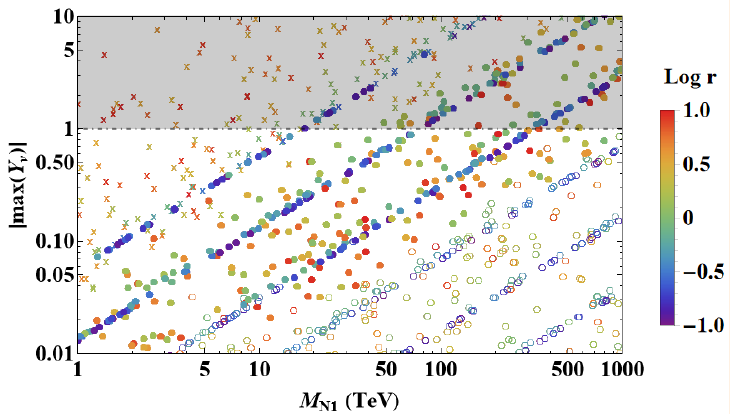}\\
    \includegraphics[width=0.48\textwidth,trim=2 2 2 2, clip]{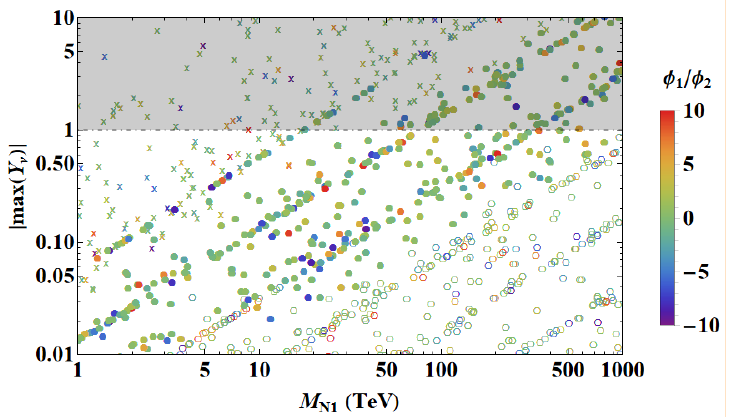}
    \includegraphics[width=0.505\textwidth,trim=2 2 2 2, clip]{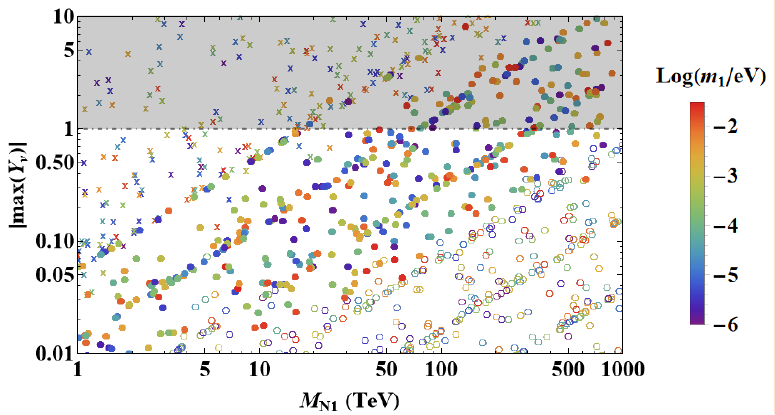}
    \caption{\em Scatter plots in the plane $\left|\max\left(Y_\nu\right)\right|$ vs. $M_{N1}$. The plots refer to the case of NO active neutrino spectrum with $\vep=0.23$. The colours indicate the value of the several parameters: top-left the PQ charge $x_\ell$, top-right $r$, bottom-left the ratio $\phi_1/\phi_2$, and bottom-right the lightest active neutrino mass $m_1$. The crosses represent points ruled out because of experimental limits, the empty circles show the points allowed but outside of the prospects, whereas the full circles correspond to points allowed within the prospects. The shaded area represents the exclusion due to the perturbativity condition. The $Log$ in the legends refers to the decimal logarithm.}
    \label{fig:CaseA_Scatter_Parameters}
\end{figure}

\begin{figure}[h!]
    \centering
    \includegraphics[width=0.49\textwidth,trim=1 1 1 1, clip]{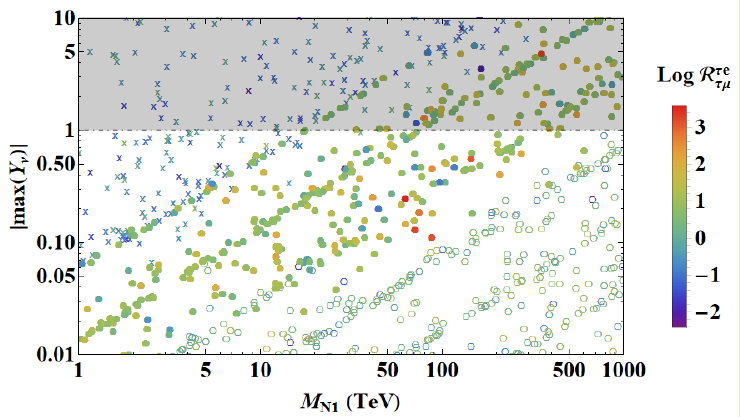}
    \includegraphics[width=0.49\textwidth,trim=1 1 1 1, clip]{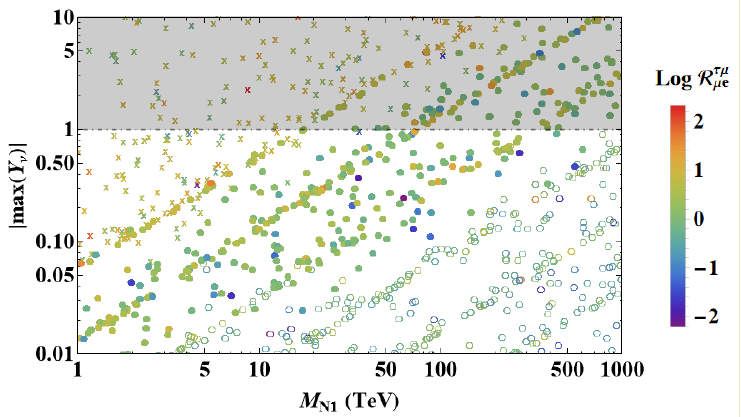}
    \caption{\em Scatter plots in the plane $\left|\max\left(Y_\nu\right)\right|$ vs. $M_{N1}$, showing in colours the value of the several parameters: left $\cR^{\tau e}_{\tau \mu}$ and right $\cR^{\tau \mu}_{\mu e}$. The plots refer to the case of NO active neutrino spectrum with $m_1=0$ and $\vep=0.23$. Symbols are the same as in Fig.~\ref{fig:CaseA_Scatter_Parameters}, while colours refer now to the observables displayed.}
    \label{fig:CaseA_Scatter_Ratios}
\end{figure}

We now explore the parameter space without imposing any relation between the three $\phi_i$. The results are shown in Figs.~\ref{fig:CaseA_Scatter_Parameters} and \ref{fig:CaseA_Scatter_Ratios}. We approach the multidimensionality of the parameter space by performing a scatter plot, generating 2000 points with $M_{N1}\in[1,10^3]\TeV$, $r\in[0.1,\,10]$, $\phi_1/\phi_2\in\pm[0.1,\,10]$, $\phi_2/\phi_3\in\pm[0.1,\,10]$, and $x_\ell$ being an integer in the range $[5,13]$. The latter range is set in order to satisfy the hierarchies in the full neutral mass matrix (and therefore the approximation under which the diagonalisation is performed, which requires $x_\ell\geq5$), and also to respect the experimental bounds or the perturbativity condition (which are not satisfied for $x_\ell>13$). 

In the plots in Fig.~\ref{fig:CaseA_Scatter_Parameters}, we show the parameter space $\left|\max\left(Y_\nu\right)\right|$ vs. $M_{N1}$, for the NO active neutrino spectrum. Crosses represent points ruled out because of experimental limits, while empty circles show the points allowed but beyond the prospects, and  full circles correspond to points allowed within the prospects. Colours refer to the intensity of the parameters considered: $x_\ell$ in the top-left plot, $r$ in the top-right, and the ratio $\phi_1/\phi_2$ in the bottom-left; in the bottom-right plot, we show the dependence on the lightest active neutrino mass. The four plots give complementary information that are well consistent with the results in Figs.~\ref{fig:CaseA_MFV} and \ref{fig:CaseA_Benchmarks}. All the plots  present higher densities of points along lines: this is the effect of the discretisation of $x_\ell$. 

In the top-left plot we can identify three main regions: top-left where most of the points are crosses; bottom-right where most of the points are empty circles; and diagonal bands of red and orange full circles. The latter correspond to $x_\ell\gtrsim 11$ for masses of the lightest heavy neutral lepton that span a pretty large window of values, from $1\TeV$ up to $400\TeV$. However, smaller values of $x_\ell$ (blue or purple full circles) are also present, but are associated to scattered values of $M_{N1}$. As Fig.~\ref{fig:CaseA_MFV} shows, the values of $\left|\max\left(Y_\nu\right)\right|$ corresponding to these points are expected to be (much) smaller than $10^{-2}$; here we can appreciate the effect of $\cH$, enhancing $\left|\max\left(Y_\nu\right)\right|$ without spoiling the description of the active neutrino masses. This is confirmed in the top-right plot, where the same points of the plot on the left are reported, but showing the dependence on $r$. The points in the diagonal bands corresponds to smaller values of $r$ (blue full circles), while those associated to larger values of $r$ are more randomly distributed. These features are due to the fact that for small values of $r$, $\cH$ tends to the unity matrix, and therefore $\left|\max\left(Y_\nu\right)\right|$ is mainly determined only by $M_{N1}$ and $x_\ell$. This relation is lost once $r$ acquires large values.

The plot in the bottom-left contains the information of the hierarchy between the $\phi_1$ and $\phi_2$ parameters. The $\phi_2/\phi_3$ plot looks very similar to this one. First of all, the lack of points associated to very large and very small values of the ratios -- red and blue points -- is simply a consequence of the chosen flat distribution of the values of $\phi_i$. Moreover, the crosses and the full and empty circles have similar densities and are similarly distributed in terms of colours: this indicates that hierarchies among $\phi_i$ do not lead to specific phenomenological features. 

The last plot, bottom-right, shows the dependence on the lightest active neutrino mass: the density of the colours is pretty uniform in the whole parameter space, indicating the absence of a strong dependence on $m_1$. The other three plots have been obtained with $m_1=0$ and the only effect of a non-vanishing mass is to slightly disperse the points. Very similar conclusions hold for the Majorana phases, and thus we do not present explicitly any dedicated plot.

\begin{figure}[h!]
    \centering
     \includegraphics[width=0.47\textwidth,trim=1 1 1 1, clip]{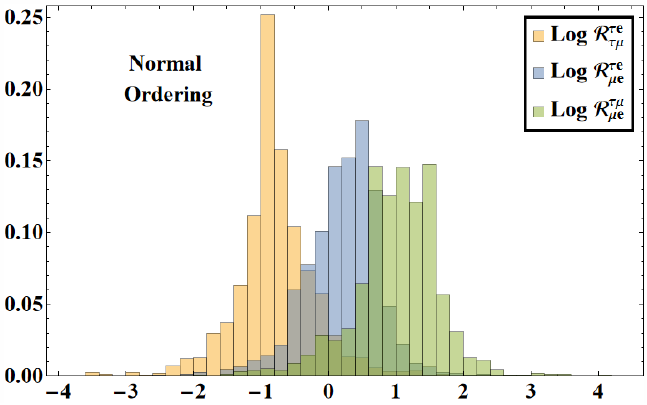}\qquad
    \includegraphics[width=0.47\textwidth,trim=1 1 1 1.25, clip]{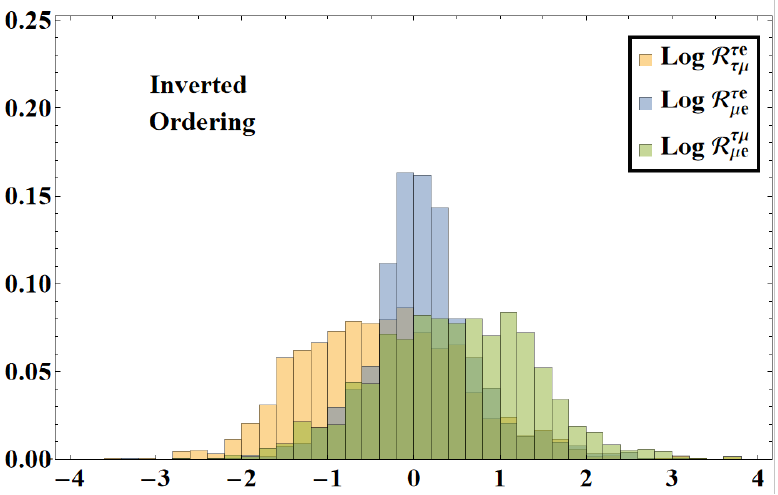}
    \caption{\em Density histograms for the ratios of the branching ratios $\cR^{\ell_i\ell_j}_{\ell_k\ell_s}$ for NO on the left and for IO on the right. In orange $\cR^{\tau e}_{\tau \mu}$, in blue $\cR^{\tau e}_{\mu e}$ and in green $\cR^{\tau \mu}_{\mu e}$. The input is the set of points allowed by the present experimental bounds and by the perturbativity condition.}
    \label{fig:CaseA_Scatter_RatiosHisto}
\end{figure}

The plots of Figs.~\ref{fig:CaseA_Scatter_Ratios} and \ref{fig:CaseA_Scatter_RatiosHisto} conclude our analysis of this scenario. The parameter space and the symbols of the points in the plots in Fig.~\ref{fig:CaseA_Scatter_Ratios} are the same as in Fig.~\ref{fig:CaseA_Scatter_Parameters}, while the colour code refers to the magnitude of the ratios among the branching ratios of the radiative rare charged lepton decays:
\be
\cR^{\ell_i\ell_j}_{\ell_k\ell_s}\equiv \dfrac{\BR(\ell_i\to\ell_j\gamma)}{\BR(\ell_k\to\ell_s\gamma)}=\dfrac{\left|\eta_{\ell_j\ell_i}\right|^2}{\left|\eta_{\ell_s\ell_k}\right|^2}\,,
\ee
where the last equality follows from Eq.~\eqref{BRmutoegamma}. The three ratios are not independent, and in particular we will present in Fig.~\ref{fig:CaseA_Scatter_Ratios} only plots for $\cR^{\tau e}_{\tau \mu}$ and $\cR^{\tau \mu}_{\mu e}$, as any conclusion on $\cR^{\tau e}_{\mu e}$ can be obtained noticing that 
\be
\cR^{\tau e}_{\mu e} = \cR^{\tau e}_{\tau \mu} \cR^{\tau \mu}_{\mu e}\,.
\ee

The two plots are very similar to each other: the upper region is excluded due to the perturbativity condition; the extreme left part of the parameter space is excluded by current bounds, while the extreme right one is beyond the future prospects; the central oblique region of the full circles are values of the parameter space that may be testable in the near future. The plots indicate an approximate direct proportionality between $M_{N1}$ and $\left|\max\left(Y_\nu\right)\right|$. Interestingly, in the two plots, the smallest $M_{N1}$ is of a few TeV and corresponds to not so small values of $\left|\max\left(Y_\nu\right)\right|$.

In each plot, there is a dominant colour, indicating a most frequent value of the ratios of the branching ratios. This can be easily seen in the left plot of Fig.~\ref{fig:CaseA_Scatter_RatiosHisto}, that represents a density histogram for the three observables: in orange $\cR^{\tau e}_{\tau \mu}$, in blue $\cR^{\tau e}_{\mu e}$ and in green $\cR^{\tau \mu}_{\mu e}$. The histogram input is the set of points in Fig.~\ref{fig:CaseA_Scatter_Ratios} that pass the present experimental bounds and the perturbativity condition, that is the full and empty circles. Each distribution, despite being not very narrow, is centred around a specific value: the ranges of the most frequent values are 
\be
\cR^{\tau e}_{\tau \mu}\sim0.1\,,\qquad\qquad
\cR^{\tau \mu}_{\mu e}\sim[3,\,30]\,,\qquad\qquad
\cR^{\tau e}_{\mu e}\sim[0.3,\,3]\,,
\ee
corresponding to a well determined hierarchy among the branching ratios of the different processes,
\be
\BR(\mu\to e\gamma)\lesssim\BR(\tau\to e\gamma)<\BR(\tau\to\mu\gamma)\,.
\ee
This is consistent with the results presented in Ref.~\cite{Dinh:2017smk} for this case (dubbed EFCI in that paper), when considering the different input parameters.

Assuming the observation of the muon radiative rare decay with a value close to the present bound, the predicted values of the branching ratios of the tau radiative rare decays are $\BR(\tau\to e\gamma)\sim 1.8\times 10^{-13}$ and $\BR(\tau\to \mu\gamma)\sim 6\times 10^{-13}$, that unfortunately are far from the present experimental bounds and will hardly be achievable in the future.

We conclude this section commenting on the IO active neutrino mass spectrum scenario. As already stated at the beginning of this section, the IO case does not lead to significantly different phenomenology with respect to the NO one. The equivalent plots in Fig.~\ref{fig:CaseA_Scatter_Parameters} for the IO scenario are very similar. Besides, the general behaviour shown in the plots in Fig.~\ref{fig:CaseA_Scatter_Ratios} is also present for the IO case, although the most frequent values of the ratios of the branching ratios are slightly different. This can be clearly appreciated looking at the density histogram in the right plot in Fig.~\ref{fig:CaseA_Scatter_RatiosHisto}. The distributions are not very peaked towards a certain value. Nevertheless, we can conclude that
\be
\cR^{\tau e}_{\tau \mu}\sim[0.03,\,1]\,,\qquad\qquad
\cR^{\tau \mu}_{\mu e}\sim[1,\,30]\,,\qquad\qquad
\cR^{\tau e}_{\mu e}\sim1\,,
\ee
which complicates the determination of a clear hierarchy among the branching ratios, besides concluding that $\BR(\tau\to\mu\gamma)$ dominates over those of the other processes. In any case, assuming a determination of the muon decay with a branching ratio equal to the present experimental bound, the corresponding predictions for the tau decays do not look hopeful for an observation of these decays in the future.

\boldmath
\section{$\cG_F^\text{NA}=SU(3)_{V}\times SU(3)_{e_R}$ with $N_R\sim S_R\sim3$}
\unboldmath
\label{sec:CASEB}

We now discuss the second scenario, CASE B, that is characterised by a smaller symmetry group with respect to the previous framework:
\be
\cG_F^{SU(3)_V}\equiv SU(3)_{V}\times SU(3)_{e_R}\times \UPQ\,,
\label{GroupSU3V}
\ee
where the first term, $SU(3)_{V}$, is the symmetry of both the EW lepton doublet and the exotic neutral leptons. Moreover, in this case, the two types of exotic neutral leptons transform identically under $SU(3)_{V}$. The corresponding transformation properties are collected in Tab.~\ref{Tab.CaseB}.

\begin{table}[h!]
\centering
\begin{tabular}{c|ccc|c} 
&  $SU(3)_{V} $ & $SU(3)_{e_R}$ & $U(1)_{PQ}$\\ 
\hline 
\hline 
&&&\\[-3mm]
$\ell_L$  	& $\bf 3$ 		& 1		    & $x_\ell$\\
$e_R$   	& 1  			& $\bf 3$   & $x_e$ \\
$N_R$  		& $\bf 3$  		& 1 	    & $x_\ell$\\
$S_R$    	& $\bf 3$  		& 1 		& $-x_\ell$\\
$\Phi$  	& 1 			& 1 		& $-1$ \\
\hline
&&&\\[-3mm]
$Y_e$   	& $\bf3$  		& $\bf\ov3$	& 0\\
$Y_N$	    & $\bf\ov6$ 	& 1  		& 0\\
\end{tabular}
\caption{\em Transformation properties of the SM leptons, exotic neutral leptons and spurions under the global symmetries $\cG_F^{SU(3)_V}$.}
\label{Tab.CaseB}
\end{table}

The mass Lagrangian invariant under this symmetry can be written as
\ba
-\sL^\text{B}_Y=&\phantom{+}\ov{\ell_L}HY_ee_R\left(\dfrac{\Phi}{\Lambda_\Phi}\right)^{x_e-x_\ell}+
c_{\nu N}\ov{\ell_L}\tH N_R+
c_{\nu S}\ov{\ell_L}\tH S_R\left(\dfrac{\Phi}{\Lambda_\Phi}\right)^{2x_\ell}+\\
&+\dfrac12c_N\ov{N_R^c}Y_N N_R\Phi\left(\dfrac{\Phi}{\Lambda_\Phi}\right)^{2x_\ell-1}+
\dfrac12\ov{S_R^c}Y_NS_R\Phi^\dag\left(\dfrac{\Phi^\dag}{\Lambda_\Phi}\right)^{2x_\ell-1}+\\
&+\Lambda\,\ov{N_R^c}Y_NS_R+\hc\,,
\ea
where $c_i$ are free real parameters and $\Lambda$ is a real scale. The only neutrino spurion now is $Y_N$, that is symmetric, $Y_N=Y_N^T$.

The neutral mass matrix in the symmetry broken phase reads
\be
\cM_\chi=
\left(
\begin{array}{ccc}
0  
& c_{\nu N} \dfrac{v}{\sqrt2}  
& c_{\nu S} \dfrac{v}{\sqrt2}\vep^{2x_\ell}  \\[4mm]
c_{\nu N} \dfrac{v}{\sqrt2}  
& c_N\dfrac{v_\Phi}{\sqrt2}\vep^{2x_\ell-1}  \,Y_N 
& \Lambda\,Y_N  \\[4mm]
c_{\nu S}\dfrac{v}{\sqrt2}\vep^{2x_\ell}  
& \Lambda\, Y_N 
& \dfrac{v_\Phi}{\sqrt2}\vep^{2x_\ell-1}  \,Y_N 
\end{array}
\right)\,.
\ee
The spontaneous breaking of the PQ symmetry guarantees the LSS hierarchical structure, and by block diagonalising the neutral mass matrix, the mass eigenvalues read at leading order
\be
m_\nu=\dfrac{v^2}{2}\vep^{2x_\ell-1}\left(\dfrac{c^2_{\nu N} v_\Phi}{\sqrt2\Lambda^2}-\dfrac{2c_{\nu N}c_{\nu S}\vep}{\Lambda}\right)Y_N^{-1}\,,\qquad\qquad
M_N\simeq \Lambda\,Y_N\,.
\label{CaseBNeuMassMatrices}
\ee
Notice that in this case, the exotic neutral leptons are not all degenerate, but the heavy neutral mass matrices contain flavour information through the dependence on $Y_N$.
The mixing $\Theta$ between the active and exotic neutral leptons is given in this case by
\be
\Theta\simeq\dfrac{c_{\nu N}v}{\sqrt2\Lambda}Y_N^{-1}\,.
\label{ThetaCaseB}
\ee

We can now proceed with the determination of the spurions exclusively in terms of lepton masses, the entries of the PMNS and the scale $f$. The charged lepton spurion follows the same relation as in the previous case, taking a background value such as in Eq.~\eqref{ChargedLeptonSpurion}. On the other hand, from Eq.~\eqref{CaseBNeuMassMatrices} and considering the same definition of the $U$ matrix as in Eq.~\eqref{DiagonalisationOfmnu}, we obtain 
\be
Y_N=f\, U^\ast \widehat{m}_\nu^{-1} U^\dag\,,
\ee
where
\be
f\equiv\dfrac{v^2}{2}\vep^{2x_\ell-1}\left(\dfrac{c_{\nu_N}^2v_\Phi}{\sqrt2\Lambda^2}-\dfrac{2c_{\nu_N}c_{\nu_S}\vep}{\Lambda}\right)\,.
\label{fDefinitionCaseB}
\ee
Differently from the previous case, the spurion is now uniquely determined in terms of known quantities and the scale $f$, guaranteeing a strong predictive power of this scenario. Indeed, the only $d=6$ operator generated at tree level  after integrating out the exotic neutral leptons is the one defined in Eq.~\eqref{Genericd6Operator}, with the associated Wilson coefficient
\be
c_{\cO_{d=6}}=\dfrac{c_{\nu N}^2}{\Lambda^2}Y_N^{-1}\left(Y_N^{-1}\right)^\dag=\dfrac{c_{\nu N}^2}{f^2\Lambda^2}
U\,\widehat{m}_\nu^2\,U^\dagger \,.
\ee

\subsection*{Phenomenological Consequences}

\begin{figure}[h!]
    \centering
    \includegraphics[width=0.41\textwidth,trim=2 2 65 2, clip]{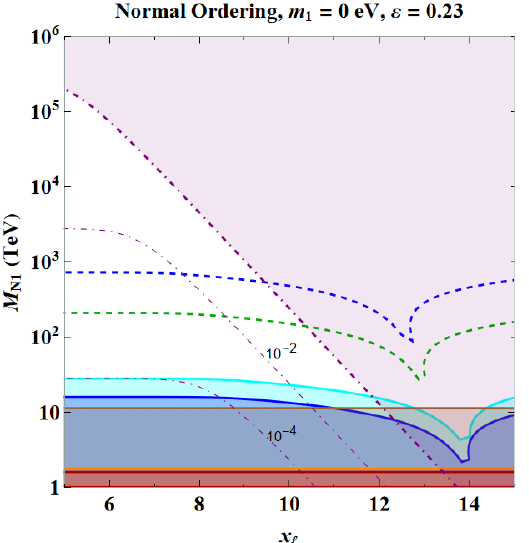}
    \includegraphics[width=0.58\textwidth,trim=2 2 2 2, clip]{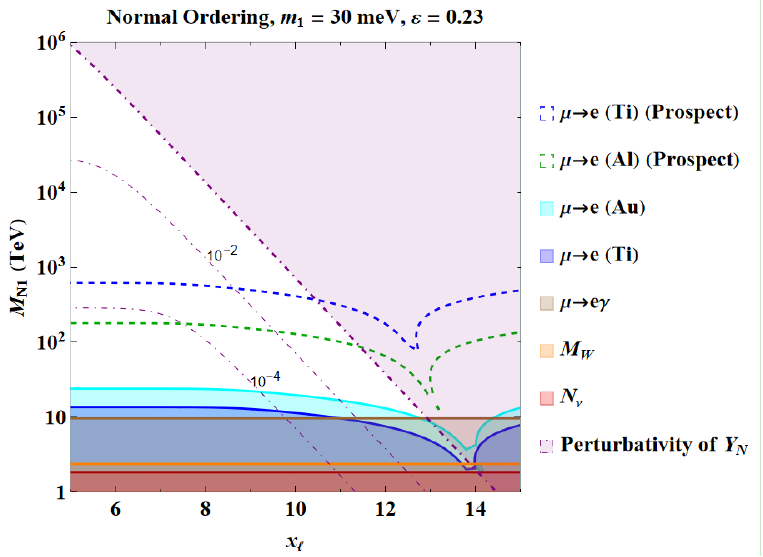}\\
     \includegraphics[width=0.41\textwidth,trim=1 1 65 1, clip]{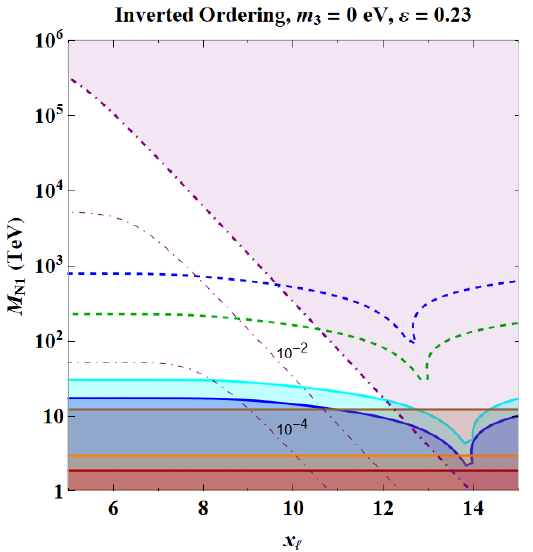}
     \includegraphics[width=0.58\textwidth,trim=1 1.5 1 1, clip]{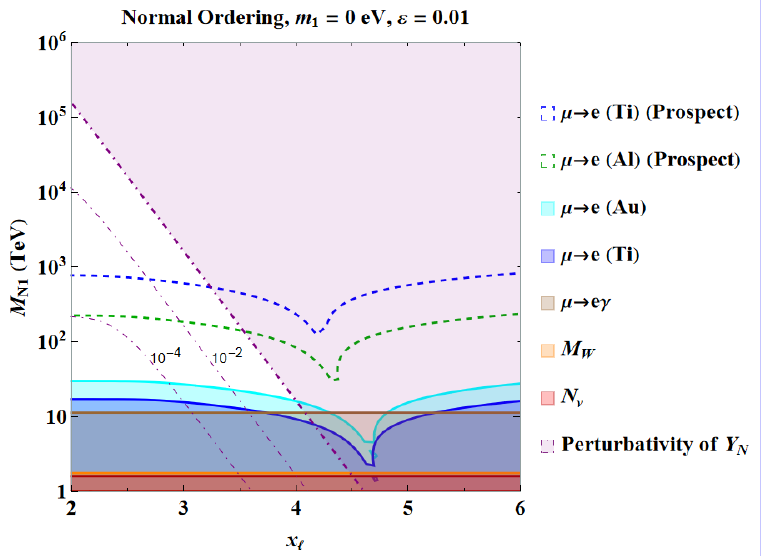}
    \caption{\em Parameter space of the lightest neutral exotic lepton mass $M_{N1}$ vs. the PQ charge $x_\ell$. Lines and colours are explained in the legend and are the same as in Figs.~\eqref{fig:CaseA_MFV} and \eqref{fig:CaseA_Benchmarks}. The plot on the top-left is for the NO case with a massless lightest active neutrino and $\vep=0.23$. The one on the top-right differs from the previous one in that $m_1=30\meV$. The one in the bottom-left corresponds to the IO case with a massless lightest active neutrino and $\vep=0.23$. The plot on the bottom-right is the only one with $\vep=0.01$ and it refers to NO with $m_1=0$.}
    \label{fig:Case2_Observables}
\end{figure}

We can now proceed with the computation of the same observables as described in the previous section. In order to do so, we first introduce the explicit expression for $\eta$ in this scenario:
\be
\eta=\dfrac{c_{\nu N}^2v^2}{4\Lambda^2}Y_N^{-1}\left(Y_N^{-1}\right)^\dag=\dfrac{c_{\nu N}^2v^2}{4f^2\Lambda^2}U\,\widehat{m}_\nu^2\,U^\dagger\,.
\ee
As already stated, the predictive power of this scenario is much higher with respect to the previous case, due to the selection of $Y_N$ as the only neutrino spurion. Moreover, the dependence on the Majorana phases $\alpha_{2,3}$ disappears in $\eta$, further reducing the parameter space for this scenario. Consistently with the previous study, we fix the order 1 parameters equal to $c_{\nu N}=c_{\nu S}=1$.  We repeat an analysis based on the same observables considered in the previous section, showing the main results in Fig.~\ref{fig:Case2_Observables}.

The plots in this figure illustrate the allowed parameter space of the lightest exotic neutral lepton mass $M_{N1}$ vs. the PQ charge $x_\ell$, once the present experimental bounds and the perturbativity condition of the neutrino spurion $Y_N$ are imposed. The future prospects for $\mu\to e$ conversion in nuclei are also shown. The characteristic cusps associated to the experimental bounds are also present in these plots. The colour code and the choice of the lines are defined in the legends consistently with the analysis in the previous section.

The four plots are very similar,  and constrain the allowed parameter space for the lightest exotic neutral lepton, $M_{N1}\sim[10,\,10^5]\TeV$, and the PQ charge, $x_\ell\lesssim13$ ($x_\ell\lesssim4$) for $\vep=0.23$ ($\vep=0.01$). For the largest allowed values of $x_\ell$, the range of values of $M_{N1}$ becomes more and more constrained to its smallest allowed value, $M_{N1}\sim10\TeV$. While the experimental bounds constrain the lower part of the parameter space, the perturbativity condition affects an almost complementary region: when requiring that $Y_N$ remains within the ``natural'' window, i.e. $\left|\max\left(Y_N\right)\right|\in[10^{-2},\,1]$, the allowed area is further reduced, implying that smaller values of $M_{N1}$ correspond only to larger values of $x_\ell$. To better understand the constraint due to the perturbativity condition, it is useful to manipulate Eq.~\eqref{CaseBNeuMassMatrices}, expressing the active neutrino mass matrix in terms of $M_N$. The dominant contribution reads
\be
m_\nu\propto \dfrac{\vep^{2x_\ell-1}}{M_N^2}Y_N\,.
\ee
It is straightforward now to understand that the top-right region of the plots is excluded by the perturbativity condition.

Despite their similarities, the four plots provide complementary information. Comparing the two plots in the top, they both refer to the NO active neutrino spectrum and with $\vep=0.23$, but the lightest neutrino mass is set to two different values: on the left $m_1=0$ while on the right $m_1=30\meV$, that is the largest  value consistent with Eq.~\eqref{LargestValuesm1m3}. The biggest difference is due to the perturbativity condition: when increasing the lightest active neutrino mass, the bound shifts to larger values of $M_{N1}$ and, although only slightly, of $x_\ell$. The experimental constraints are less sensitive to $m_1$.

The plot in the bottom-left shows the IO case, for a massless lightest active neutrino and $\vep=0.23$. By comparison with its NO counterpart, we conclude that also in this scenario the ordering plays a minor role. 

The last plot, in the bottom-right, is for $\vep=0.01$, NO and with $m_1=0$. The main effect of changing the value of $\vep$ is, as expected, a shift in the values of $x_\ell$, since the dependence is through $\vep^{2 x_\ell-1}$, as mentioned in the previous section. On the other hand, comparing it with the plot in the top-left, we can see that the perturbativity condition is stronger: if we require $Y_N$ to be in the ``natural'' regime, a smaller region of the allowed parameter space is left, indicating an inverse proportionality between $M_{N1}$ and $x_\ell$.

Comparing the results of this CASE B with the ones of the previous CASE A, we can notice that in this scenario the exotic neutral leptons cannot be as light as in the previous scenario, excluding any possibility of direct production and detection of these states at colliders. For what concerns the values of $x_\ell$, the allowed range of values is very similar.

On the other hand, indirect searches can be much more promising in providing information about this new physics. In this framework, present experimental bounds on $\mu\to e\gamma$ can test values of the NP scale up to $\sim10\TeV$, while the future prospects on the $\mu\to e$ conversion in nuclei may reach up to $10^{3}\TeV$. 

A complementary test is provided by the ratios of branching ratios of radiative rare decays of charged leptons, $\cR^{\ell_i\ell_j}_{\ell_k\ell_s}$. This scenario is so much more predictive than the previous one that we can identify specific values of these observables for the NO (IO) active neutrino spectrum:
\be
\cR^{\tau e}_{\mu e}=1.5\,(0.8)\,,\qquad\qquad
\cR^{\tau e}_{\tau \mu}=0.06\,(0.04)\,,\qquad\qquad
\cR^{\tau \mu}_{\mu e}=27\,(20)\,.
\ee
We can therefore identify clear hierarchies between the branching ratios of the different processes:
\ba
\BR(\mu\to e\gamma)<\BR(\tau\to e\gamma)<\BR(\tau\to\mu\gamma)&\qquad\text{Normal Ordering,}\\[2mm]
\BR(\tau\to e\gamma)\lesssim\BR(\mu\to e\gamma)<\BR(\tau\to\mu\gamma)&\qquad\text{Inverted Ordering}\,.
\ea
Unfortunately, as in the previous case, the predicted values of the tau decay branching ratios, assuming the observation of the muon decay, are extremely small.

We conclude this section comparing these results with those obtained in Ref.~\cite{Dinh:2017smk}, and in particular to the case dubbed EFCII. For the NO case, a strong dependence on the lightest neutrino mass had been pointed out in the $\cR^{\ell_i\ell_j}_{\ell_k\ell_s}$ observables. This apparent contradiction finds an explanation in the different spurion analysis performed in the two studies: we explicitly integrated out the heavy neutral leptons obtaining that $\eta\propto Y_N^{-1}\left(Y_N^{-1}\right)^\dag$; in Ref.~\cite{Dinh:2017smk}, instead, $\eta$ had been computed with the traditional spurion insertion technique, obtaining that $\eta\propto Y_N^\dag Y_N$ (see $\Delta$ in Eq.~(2.29) in Ref.~\cite{Dinh:2017smk}). As a result, while the dependence on $m_1$ is negligible in this analysis, it is instead dominant in the study presented in Ref.~\cite{Dinh:2017smk}. In the IO case, both analyses agree: despite the different definition of $\eta$, the dependence on $m_3$ is negligible and the predicted ratios of branching ratios are very similar.

\boldmath
\section{$\cG_F^\text{NA}=SU(3)_{V}\times SU(3)_{e_R}$ with $N_R\sim \ov{S_R}\sim3$}
\unboldmath
\label{sec:CASEC}

The last scenario that we will investigate is based on the same flavour symmetry group as the former case, $\cG_F^{SU(3)_V}$, as defined in Eq.~\eqref{GroupSU3V}. The main difference in the model building are the transformation properties of the exotic neutral leptons: while $N_R$ transforms as a triplet, $S_R$ does so as an anti-triplet of the $SU(3)_V$ group. The transformation properties under $\cG_F^{SU(3)_V}$ are collected in Tab.~\ref{Tab.CaseC}.

\begin{table}[h!]
\centering
\begin{tabular}{c|ccc|c} 
&  $SU(3)_{V} $ & $SU(3)_{e_R}$ & $U(1)_{PQ}$\\ 
\hline 
\hline 
&&&\\[-3mm]
$\ell_L$  	& $\bf 3$ 		& 1		    & $x_\ell$\\
$e_R$   	& 1  			& $\bf 3$   & $x_e$ \\
$N_R$  		& $\bf 3$  		& 1 	    & $x_\ell$\\
$S_R$    	& $\bf \ov3$  		& 1 		& $-x_\ell$\\
$\Phi$  	& 1 			& 1 		& $-1$ \\
\hline
&&&\\[-3mm]
$Y_e$   	& $\bf3$  		& $\bf\ov3$	& 0\\
$Y_N$	    & $\bf\ov6$ 	& 1  		& 0\\
\end{tabular}
\caption{\em Transformation properties of the SM leptons, exotic neutral leptons and spurions under the global symmetries $\cG_F^{SU(3)_V}$.}
\label{Tab.CaseC}
\end{table}

The mass Lagrangian invariant under this symmetry can be written as
\ba
-\sL^\text{C}_Y=&\phantom{+}\ov{\ell_L}HY_ee_R\left(\dfrac{\Phi}{\Lambda_\Phi}\right)^{x_e-x_\ell}+
c_{\nu N}\ov{\ell_L}\tH N_R+
c_{\nu S}\ov{\ell_L}\tH\,Y_N^\dag S_R\left(\dfrac{\Phi}{\Lambda_\Phi}\right)^{2x_\ell}+\\
&+\dfrac12c_N\ov{N_R^c}Y_N N_R\Phi\left(\dfrac{\Phi}{\Lambda_\Phi}\right)^{2x_\ell-1}+
\dfrac12\ov{S_R^c}Y_N^\dag S_R\Phi^\dag\left(\dfrac{\Phi^\dag}{\Lambda_\Phi}\right)^{2x_\ell-1}+\\
&+\Lambda\,\ov{N_R^c}S_R+\hc\,,
\ea
where $c_i$ are free real parameters and $\Lambda$ is a real scale. As for  CASE B, the only neutrino spurion is the symmetric $Y_N$. Notice the differences with respect to $\sL^\text{B}_Y$: the flavour information now appears both in the Majorana and Dirac mass terms.

The neutral mass matrix in the symmetry broken phase reads
\be
\cM_\chi=
\left(
\begin{array}{ccc}
0  
& c_{\nu N} \dfrac{v}{\sqrt2}  
& c_{\nu S} \dfrac{v}{\sqrt2}\vep^{2x_\ell} \,Y_N^\dag \\[4mm]
c_{\nu N} \dfrac{v}{\sqrt2} 
& c_N\dfrac{v_\Phi}{\sqrt2}\vep^{2x_\ell-1}  \,Y_N 
& \Lambda  \\[4mm]
c_{\nu S}\dfrac{v}{\sqrt2}\vep^{2x_\ell}\,Y_N^\dag
& \Lambda 
& \dfrac{v_\Phi}{\sqrt2}\vep^{2x_\ell-1}  \,Y_N^\dag 
\end{array}
\right)\,.
\ee 
The difference with respect to the previous case in the position of the neutrino spurion $Y_N$ leads to slightly different expressions for the neutrino masses:
\be
m_\nu=\dfrac{v^2}{2}\vep^{2x_\ell-1}\left(\dfrac{c^2_{\nu N} v_\Phi}{\sqrt2\Lambda^2}-\dfrac{2c_{\nu N}c_{\nu S}\vep}{\Lambda}\right)Y_N^\dag\,,\qquad\qquad
M_N\simeq \Lambda\,.
\label{CaseCNeuMassMatrices}
\ee
In this case, the six heavy neutral leptons are exactly degenerate at this level of approximation. 

The mixing $\Theta$ between the active and exotic neutral leptons is given in this case by
\be
\Theta\simeq\dfrac{c_{\nu N}v}{\sqrt2\Lambda}\,,
\label{ThetaCaseC}
\ee
that is independent on any flavour information. This is different than what occurs in the previous two scenarios, and leads to sensibly different phenomenological results. 

Regarding the determination of the spurions in terms of lepton masses and PMNS mixing, the charged lepton spurion satisfies again the relation in Eq.~\eqref{ChargedLeptonSpurion}, while for the neutrino one we obtain
\be
Y_N=\dfrac1f\, U^\ast \widehat{m}_\nu\,U^\dag\,,
\ee
where $f$ has the same expression as in Eq.~\eqref{fDefinitionCaseB}. Although the spurions are uniquely determined in terms of lepton masses and mixing and the scale $f$, the low-energy phenomenology does not contain any flavour information, as the coefficient of the $d=6$ operator simply reads
\be
c_{\cO_{d=6}}=\dfrac{c_{\nu N}^2}{\Lambda^2}\,.
\ee

\subsection*{Phenomenological Consequences}

The phenomenological analysis of this scenario is much simpler with respect to the previous contexts, as the low-energy observables do not depend on neither $x_\ell$ nor on the neutrino oscillation parameters. Indeed, the $\eta$ matrix now is proportional to the identity matrix,
\be
\eta=\dfrac{c_{\nu N}^2v^2}{4\Lambda^2}\,\unity\,,
\label{CASECeta}
\ee
depending on only one effective parameter $c_{\nu N}/\Lambda$. Consistently with the analysis in the previous sections, we fix the order 1 coefficient to be exactly $c_{\nu N}=1$ and discuss the effects of NP only in terms of $\Lambda$, that is in first approximation the mass of the heavy neutral leptons. 

\begin{figure}[h!]
    \centering
    \includegraphics[width=0.485\textwidth,trim=1 1 1 1, clip]{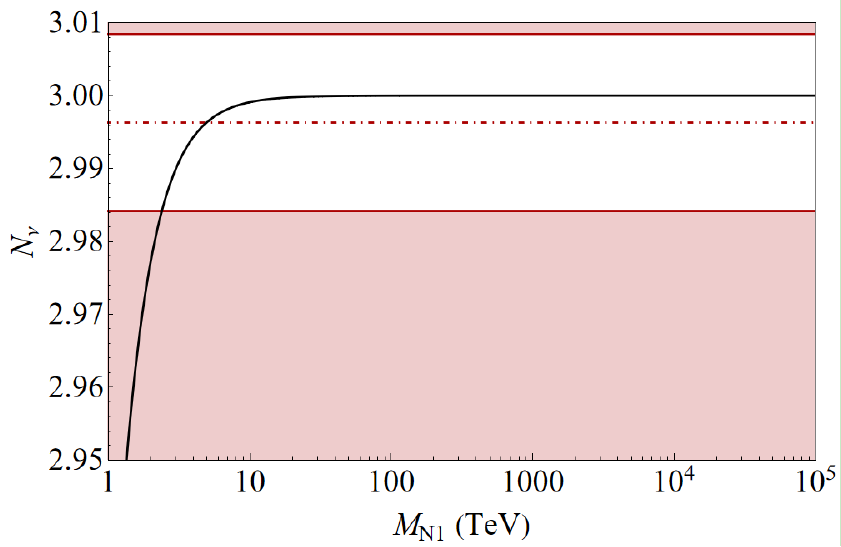}
    \includegraphics[width=0.505\textwidth,trim=1 1 1 1, clip]{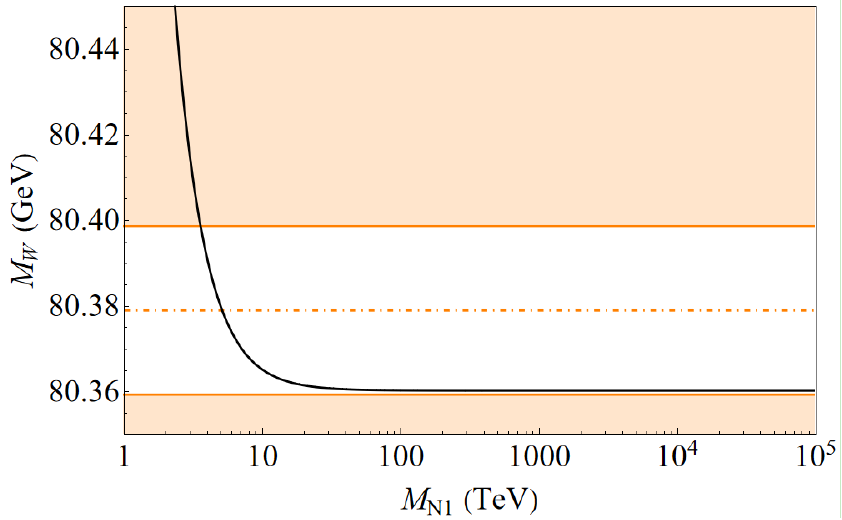}
    \caption{\em Predictions for the effective number of active neutrinos $N_\nu$ (left) and the $W$ gauge boson mass (right) as a function of the lightest heavy neutral lepton mass $M_{N1}$. The best fit values (dotted-dashed line), and the $90\%$ C.L. (shaded area) experimental limits for the observables are displayed.}
    \label{fig:Case3_Observables}
\end{figure}

The parameter $\eta$ is flavour blind and therefore in this case no deviations with respect to the SM values are predicted for the flavour violating observables discussed in the previous sections (radiative rare charged lepton decays and $\mu\to e$ conversion in nuclei). On the other hand, deviations from unitarity of the PMNS matrix $\cN$ are expected, affecting the number of active neutrinos $N_\nu$ as determined from the invisible width of the $Z$ and the $W$ boson mass through the modified decay of the muon, from which $G_F$ is extracted. This is illustrated in the plots in Fig.~\ref{fig:Case3_Observables}.

The two plots provide a very similar information: consistency with the experimental determination of the two observables is obtained with masses of the heavy neutral leptons above a few TeV, that is $M_{N1}\gtrsim2.4\TeV$ from the left plot on $N_\nu$ and $M_{N1}\gtrsim 3.6\TeV$ from the right plot on $M_W$. Although these values are still above the present sensitivity to these exotic fermions at colliders, there is hope for discovery in future facilities. 

\begin{figure}[h!]
    \centering
    \includegraphics[width=0.46\textwidth,trim=1 2 30 1, clip]{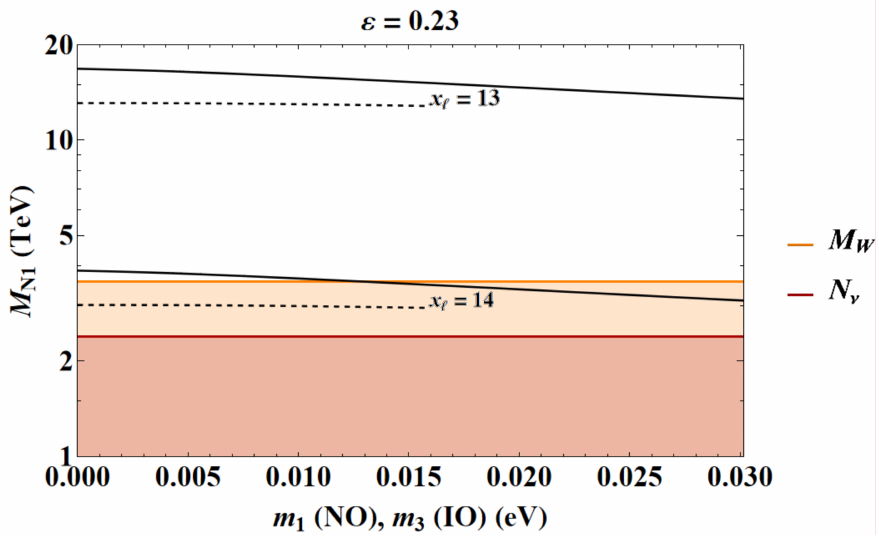}
    \includegraphics[width=0.53\textwidth,trim=1 2 2 1, clip]{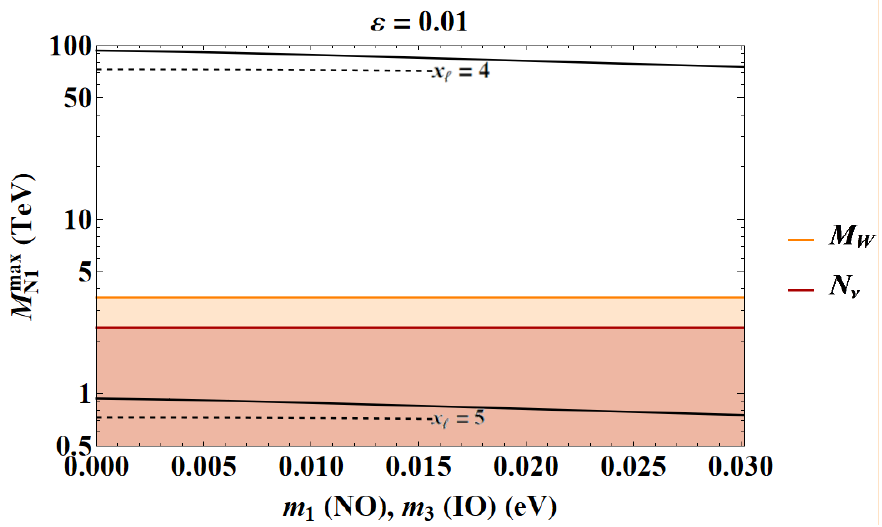}
    \caption{\em Heavy neutral lepton mass $M_{N1}$ as a function of the lightest active neutrino mass, $m_1$ for NO and $m_3$ for IO, represented for several choices of the PQ charge $x_\ell$, and for $\vep=0.23$ and $\vep=0.01$. The neutrino spurion satisfies $\left|\max\left(Y_N\right)\right|=1$. In the two top plots ($\vep=0.23$ in the left and $\vep=0.01$ in the right), solid (dashed) lines indicate NO (IO): both lines are represented in the whole range available for the lightest neutrino mass. The shaded regions refer to the experimental bounds on $M_W$ and $N_\nu$ as from the right plot in Fig.~\ref{fig:Case3_Observables}.}
    \label{fig:Case3_Observables2}
\end{figure}

With this information at hand, we can now deduce the preferred values of $x_\ell$. In Fig.~\ref{fig:Case3_Observables2} we study the dependence of the mass of the lightest heavy neutral lepton on the lightest active neutrino mass, for $\vep=0.23$ ($\vep=0.01$) on the left (right). Continuous lines refer to the NO active neutrino spectrum, and dashed lines to the IO one. The range of values of the lightest active neutrino masses is consistent with the cosmological bound in Eq.~\eqref{SumNuMasses}.

To simplify the analysis, we fix the largest absolute value of $Y_N$ at its maximal value allowed by the perturbativity condition, $\left|\max\left(Y_N\right)\right|=1$. We will comment below the consequences of taking smaller values.

For $\vep=0.23$, we can see that $x_\ell\lesssim14$ to agree at $90\%$ C.L. with the bounds from $M_W$. This is the most appealing scenario, since it corresponds to the smallest masses, down to $\sim4\TeV$, allowed for the heavy neutral leptons. Notice that, in order to keep $Y_N$ ``natural'', that is order 1, smaller values of $x_\ell$ require heavier $M_{N_1}$ so as to reproduce the correct neutrino masses.

As discussed in the previous scenarios, CASE A and CASE B, choosing $\vep=0.01$ corresponds to a rescaling of the PQ charge: $x_\ell<5$. In contrast with the left plot, the masses of the heavy neutral leptons are much higher, larger than $50\TeV$, once $x_\ell$ takes the largest possible value consistent with data.

Considering the neutrino spurion within its ``natural'' range of values, the plots in Fig.~\ref{fig:Case3_Observables2} would be slightly affected. In particular, taking $\left|\max\left(Y_N\right)\right|=10^{-2}$, it is possible to further restrict the values of the PQ charge, that, in this case, would have to satisfy $x_\ell<13$ for the $\vep=0.23$ case. On the other hand, for $\vep=0.01$, $x_\ell=4$ is still the largest value allowed by the experimental bounds on $M_W$. On the other hand, for fixed values of $x_\ell$, considering smaller values of $\left|\max\left(Y_N\right)\right|$ allows smaller values of $M_{N1}$, being an interesting possibility for detection at colliders.

\boldmath
\section{Implications of the recent measurement of $M_W$}
\unboldmath
\label{sec:MWnew}

The very new determination of the $W$ gauge boson mass at the CDF II detector~\cite{CDF:2022hxs},
\be
M_W=80.4335(94)\GeV\,,
\label{NewMWValue}
\ee
is in significant tension with the SM prediction. When taking this value into account in our analysis, the parameter space is strongly affected. Indeed, in CASEs A and B, if we attempt to solve the new tension in $M_W$, the corresponding parameter space is excluded by the lepton flavour violating processes. In particular, in CASE A, the $M_W$ dependence on $M_{N1}$ is very mild for the whole range considered, that is $M_{N1}\in[1,\,10^6]\TeV$, while it does depend on $x_\ell$. For the specific NO benchmark with $m_1=0$, $\vep=0.23$ and $r=0$, a value of $M_W$ within the 90\% C.L. of the new measurement is obtained for $x_\ell=15$, which is excluded from both the $N_\nu$ and all the lepton flavour violating processes. Allowing for a non-vanishing $m_1$, lowering $\vep$ or considering the IO scenario does not lead to a different conclusion. A non-vanishing $r$, however, may have interesting results. Indeed, to solve the $M_W$ tension, a relatively large value of $\eta_{ee}+\eta_{\mu\mu}$ is necessary, and $r>1$ increases by construction the value of the elements of $\eta$. In spite of that, it turns out that the required values of $r$ are ruled out by the other observables. For example, considering $M_{N1}=1\TeV$ and $x_\ell=13$, that is an allowed point in the left plot in Fig.~\ref{fig:CaseA_MFV} and within the future prospects for $\mu\to e$ conversion, the required value would be $r=4.6$, which is ruled out. Instead, selecting a benchmark consistent with Fig.~\ref{fig:CaseA_Benchmarks}, with $M_{N1}=10^3\TeV$ and $x_\ell=5$, the corresponding value would be $r=14$, also excluded. We can conclude that no points in the parameter space of CASE A would be consistent with the explanation of the new measurement of $M_W$ and with the experimental bounds considered.

A very similar conclusion holds for CASE B. In particular, for NO with $m_1=0$ and $\vep=0.23$, the $M_W$ tension is alleviated with $M_{N1}\in[1.1,\,1.4]\TeV$ and for $x_\ell$ in the whole allowed range. As shown in the top-left plot in Fig.~\ref{fig:Case2_Observables}, these values are excluded not only by the radiative rare muon decay and by $\mu\to e$ conversion, but also by the number of active neutrinos, which requires $M_{N1}>1.6\TeV$. Allowing for a non-vanishing lightest active neutrino mass or a smaller $\vep$ does not alter this conclusion. Moreover, considering the IO case, the favoured values for the lightest heavy neutral leptons are $M_{N1}\in[2,\,2.4]\TeV$. This range would be consistent with the bound from $N_\nu$, which in this case requires $M_{N1}>1.9\TeV$, but is excluded by the flavour violating processes. 

\begin{figure}[h!]
    \centering
    \includegraphics[width=0.55\textwidth,trim=1 0 0 0, clip]{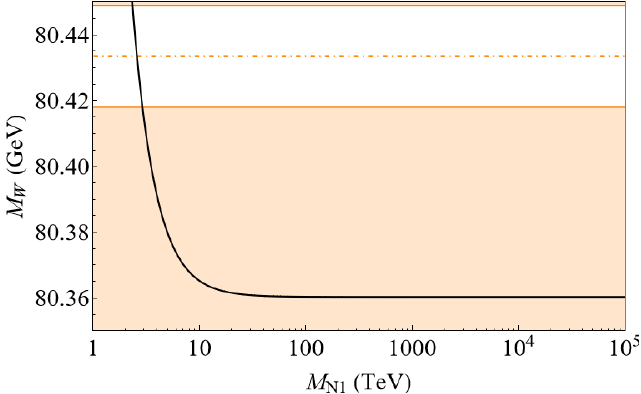}
    \caption{\em Predictions for the $W$ gauge boson mass as a function of the lightest heavy neutral lepton mass $M_{N1}$. The best fit value (dotted-dashed line) as reported in Eq.~\eqref{NewMWValue}, and the corresponding $90\%$ C.L. (shaded area) experimental limits are displayed.}
    \label{fig:Case3_Observables3}
\end{figure}

Finally, CASE C is particularly interesting in this respect, as it predicts the absence of any flavour violating process. The equivalent plot of that on the right in Fig.~\ref{fig:Case3_Observables} is reported in Fig.~\ref{fig:Case3_Observables3}. The allowed values for $M_{N1}$ considering the $90\%$ C.L. of the new experimental determination shrink to a very narrow range,
\be
M_{N1}\in[2.3,\,2.9]\TeV\,,
\label{MN1RangeSharp}
\ee
where only the lower values are excluded by the effective number of neutrinos, which puts a lower bound on $M_{N1}$ of $2.4\TeV$. 

In contrast with the other cases, this last scenario does provide an explanation of the $M_W$ tension, being consistent with the other experimental bounds and the perturbativity condition, yielding a very sharp prediction for the lightest heavy neutral lepton.
Considering further constraints on this interpretation of the new $M_W$ measurement, it is interesting to point out that lepton flavour universality tests from various meson and tau decays, that are also strong probes of the PMNS unitarity, do not apply in this CASE C, as the $\eta$ matrix in Eq.~\eqref{CASECeta} is flavour blind. On the other hand, translating the range of values of $M_{N1}$ in Eq.~\eqref{MN1RangeSharp} in terms of $\eta$, we obtain that $\eta\sim0.002$. This would worsen the present Cabibbo anomaly, which would instead require a negative value for $\eta$ (for a recent review, see Ref.~\cite{Bryman:2021teu} and references therein).

\section{Conclusions}
\label{sec:Concls}

The Inverse Seesaw mechanism is well known to display the interesting feature of separating the scale of lepton number violation from the mass scale of the exotic neutral leptons that enrich the spectrum. As a result, the associated phenomenology is not suppressed by the tiny masses of the active neutrinos, and the heavy neutral leptons could be, in general, sufficiently light to be produced and detected at colliders. In this paper we investigated a possible dynamical origin of the lepton number breaking within the Minimal Lepton Flavour Violation setup. Indeed, one of the Abelian terms of the whole flavour symmetry corresponds to the Peccei-Quinn symmetry and can also be merged with lepton number. Thus, it provides a natural mechanism to explain the smallness of the lepton number breaking parameters.

An interesting consequence is the appearance of an axion that provides a solution to the Strong CP problem and represents a possible candidate for Dark Matter, with a characteristic scale that can be as low as $f_a\sim10^{8}\GeV$. In this context, this axion plays also the role of a Majoron. While its phenomenology has been previously studied in Ref.~\cite{Arias-Aragon:2017eww}, we focused here in the effects of the new exotic neutral leptons. In particular, we studied the impact of the non-unitarity of the PMNS matrix, that can impact precision electroweak and flavour observables. Stringent constraints can be derived from the effective number of active neutrinos as determined from the invisible decay width of the $Z$, the value of the $W$ gauge boson mass when compared to its prediction from the Fermi constant from muon dominant decay and various lepton flavour violating processes, such as $\mu\to e$ conversion in nuclei and radiative rare charged lepton decays.

Notice that the Minimal Flavour Violation ansatz further sharpens the predictivity of these scenarios by pinpointing the observable which is most sensitive to new physics, given the correlations between the different coefficients of the $d=6$ operators that are derived from each specific case. We have, for the first time, explored systematically all realisations of Minimal Lepton Flavour Violation in this Low-Scale Seesaw context, identifying three different scenarios in terms of possible lepton flavour symmetries and of the transformation properties of the exotic neutral leptons. Indeed, while the most general scenario lacks predictive power due to the presence of several independent spurions, we reduce their number to two in each considered framework. We find significant differences between the three scenarios.
\begin{description}
\item[Indirect Searches:]
 CASEs A and B predict non-vanishing contributions to $\mu\to e$ conversion in nuclei and radiative rare charged lepton decays, which provide the strongest experimental bounds. The $\tau\to\mu\gamma$ decay is predicted in both cases to be the dominant one, while there is no clear hierarchy between the branching ratios of $\mu\to e\gamma$ and $\tau\to e\gamma$, once considering the different options of NO and IO. Instead, in CASE C no lepton flavour violating effects are expected, and therefore
 its best probes are electroweak precision observables.
\item[Direct Searches:]
Regarding the masses of the heavy neutral leptons, CASEs A and C allow values as low as a few TeV, while in CASE B the lightest values are about $10\TeV$. As a consequence, we may hope for a detection of the heavy neutral leptons in the future colliders only in the first two  scenarios mentioned, that is CASEs A and C.
\end{description}

In summary, the three frameworks described are therefore consistent with present experimental data from both flavour facilities and colliders. An observation of radiative rare decays or $\mu\to e$ conversion in nuclei would not be consistent with the predictions in CASE C, while it would correspond to well defined points in the parameter space for the other two cases, even if other experimental inputs will be necessary to identify a specific scenario. Besides, direct collider searches may be very promising to probe CASEs A and C.

In the light of the very recent measurement of the $W$ boson mass at the CDF II detector and its corresponding tension with the Standard Model, CASEs A and B would not be consistent with an explanation of this anomaly. On the other hand, within CASE C, a value for $M_W$ within its $90\%$ C.L. pinpoints a narrow range of values for the masses of the lightest heavy neutral leptons, at about $2-3\TeV$. Although this sharp prediction would worsen the present Cabibbo anomaly, it is consistent with the constraint given by the effective number of active neutrinos and could be testable through collider searches in the near future.

\section*{Acknowledgements}
The authors thanks Andrea Caputo, Ferruccio Feruglio and Pilar Hernandez for discussions during the initial phase of this project. E.F.M., M.G.L. and L.M. acknowledge partial financial support by the Spanish Research Agency (Agencia Estatal de Investigaci\'on) through the grant IFT Centro de Excelencia Severo Ochoa No CEX2020-001007-S and by the grant PID2019-108892RB-I00 funded by MCIN/AEI/ 10.13039/501100011033, by the European Union's Horizon 2020 research and innovation programme under the Marie Sk\l odowska-Curie grant agreement No 860881-HIDDeN. F.A.A. acknowledges funding from the French Programme d'investissements d'avenir through the Enigmass Labex and support from the IN2P3 Master project ``Axions from Particle Physics to Cosmology''.


\footnotesize

\bibliography{biblio}{}

\providecommand{\href}[2]{#2}\begingroup\raggedright\begin{thebibliography}{100}

\bibitem{Weinberg:1979sa}
S.~Weinberg, {\it {Baryon and Lepton Nonconserving Processes}},  Phys. Rev.
  Lett. {\bf 43} (1979) 1566--1570.

\bibitem{Wyler:1982dd}
D.~Wyler and L.~Wolfenstein, {\it {Massless Neutrinos in Left-Right Symmetric
  Models}},  Nucl. Phys. B {\bf 218} (1983) 205--214.

\bibitem{Mohapatra:1986bd}
R.~N. Mohapatra and J.~W.~F. Valle, {\it {Neutrino Mass and Baryon Number
  Nonconservation in Superstring Models}},  Phys. Rev. D {\bf 34} (1986) 1642.

\bibitem{Bernabeu:1987gr}
J.~Bernabeu, A.~Santamaria, J.~Vidal, A.~Mendez, and J.~W.~F. Valle, {\it
  {Lepton Flavor Nonconservation at High-Energies in a Superstring Inspired
  Standard Model}},  Phys. Lett. B {\bf 187} (1987) 303--308.

\bibitem{Broncano:2002rw}
A.~Broncano, M.~B. Gavela, and E.~E. Jenkins, {\it {The Effective Lagrangian
  for the seesaw model of neutrino mass and leptogenesis}},  Phys. Lett. B {\bf
  552} (2003) 177--184, [\href{http://arxiv.org/abs/hep-ph/0210271}{{\tt
  hep-ph/0210271}}]. [Erratum: Phys.Lett.B 636, 332 (2006)].

\bibitem{Kersten:2007vk}
J.~Kersten and A.~Y. Smirnov, {\it {Right-Handed Neutrinos at CERN LHC and the
  Mechanism of Neutrino Mass Generation}},  Phys. Rev. D {\bf 76} (2007)
  073005, [\href{http://arxiv.org/abs/0705.3221}{{\tt arXiv:0705.3221}}].

\bibitem{Abada:2007ux}
A.~Abada, C.~Biggio, F.~Bonnet, M.~B. Gavela, and T.~Hambye, {\it {Low energy
  effects of neutrino masses}},  JHEP {\bf 12} (2007) 061,
  [\href{http://arxiv.org/abs/0707.4058}{{\tt arXiv:0707.4058}}].

\bibitem{Gavela:2009cd}
M.~B. Gavela, T.~Hambye, D.~Hernandez, and P.~Hernandez, {\it {Minimal Flavour
  Seesaw Models}},  JHEP {\bf 09} (2009) 038,
  [\href{http://arxiv.org/abs/0906.1461}{{\tt arXiv:0906.1461}}].

\bibitem{Chivukula:1987py}
R.~S. Chivukula and H.~Georgi, {\it {Composite Technicolor Standard Model}},
  Phys. Lett. B {\bf 188} (1987) 99--104.

\bibitem{DAmbrosio:2002vsn}
G.~D'Ambrosio, G.~F. Giudice, G.~Isidori, and A.~Strumia, {\it {Minimal flavor
  violation: An Effective field theory approach}},  Nucl. Phys. B {\bf 645}
  (2002) 155--187, [\href{http://arxiv.org/abs/hep-ph/0207036}{{\tt
  hep-ph/0207036}}].

\bibitem{Cirigliano:2005ck}
V.~Cirigliano, B.~Grinstein, G.~Isidori, and M.~B. Wise, {\it {Minimal flavor
  violation in the lepton sector}},  Nucl. Phys. B {\bf 728} (2005) 121--134,
  [\href{http://arxiv.org/abs/hep-ph/0507001}{{\tt hep-ph/0507001}}].

\bibitem{Davidson:2006bd}
S.~Davidson and F.~Palorini, {\it {Various definitions of Minimal Flavour
  Violation for Leptons}},  Phys. Lett. B {\bf 642} (2006) 72--80,
  [\href{http://arxiv.org/abs/hep-ph/0607329}{{\tt hep-ph/0607329}}].

\bibitem{Alonso:2011jd}
R.~Alonso, G.~Isidori, L.~Merlo, L.~A. Munoz, and E.~Nardi, {\it {Minimal
  flavour violation extensions of the seesaw}},  JHEP {\bf 06} (2011) 037,
  [\href{http://arxiv.org/abs/1103.5461}{{\tt arXiv:1103.5461}}].

\bibitem{Minkowski:1977sc}
P.~Minkowski, {\it {$\mu \to e\gamma$ at a Rate of One Out of $10^{9}$ Muon
  Decays?}},  Phys. Lett. B {\bf 67} (1977) 421--428.

\bibitem{Gell-Mann:1979vob}
M.~Gell-Mann, P.~Ramond, and R.~Slansky, {\it {Complex Spinors and Unified
  Theories}},  Conf. Proc. C {\bf 790927} (1979) 315--321,
  [\href{http://arxiv.org/abs/1306.4669}{{\tt arXiv:1306.4669}}].

\bibitem{Yanagida:1979as}
T.~Yanagida, {\it {Horizontal gauge symmetry and masses of neutrinos}},  Conf.
  Proc. C {\bf 7902131} (1979) 95--99.

\bibitem{Mohapatra:1979ia}
R.~N. Mohapatra and G.~Senjanovic, {\it {Neutrino Mass and Spontaneous Parity
  Nonconservation}},  Phys. Rev. Lett. {\bf 44} (1980) 912.

\bibitem{Kagan:2009bn}
A.~L. Kagan, G.~Perez, T.~Volansky, and J.~Zupan, {\it {General Minimal Flavor
  Violation}},  Phys. Rev. D {\bf 80} (2009) 076002,
  [\href{http://arxiv.org/abs/0903.1794}{{\tt arXiv:0903.1794}}].

\bibitem{Isidori:2010kg}
G.~Isidori, Y.~Nir, and G.~Perez, {\it {Flavor Physics Constraints for Physics
  Beyond the Standard Model}},  Ann. Rev. Nucl. Part. Sci. {\bf 60} (2010) 355,
  [\href{http://arxiv.org/abs/1002.0900}{{\tt arXiv:1002.0900}}].

\bibitem{EuropeanStrategyforParticlePhysicsPreparatoryGroup:2019qin}
R.~K. Ellis {\em et.~al.}, {\it {Physics Briefing Book}: {Input for the
  European Strategy for Particle Physics Update 2020}},
  \href{http://arxiv.org/abs/1910.11775}{{\tt arXiv:1910.11775}}.

\bibitem{Cirigliano:2006su}
V.~Cirigliano and B.~Grinstein, {\it {Phenomenology of minimal lepton flavor
  violation}},  Nucl. Phys. B {\bf 752} (2006) 18--39,
  [\href{http://arxiv.org/abs/hep-ph/0601111}{{\tt hep-ph/0601111}}].

\bibitem{Grinstein:2006cg}
B.~Grinstein, V.~Cirigliano, G.~Isidori, and M.~B. Wise, {\it {Grand
  Unification and the Principle of Minimal Flavor Violation}},  Nucl. Phys. B
  {\bf 763} (2007) 35--48, [\href{http://arxiv.org/abs/hep-ph/0608123}{{\tt
  hep-ph/0608123}}].

\bibitem{Paradisi:2009ey}
P.~Paradisi and D.~M. Straub, {\it {The SUSY CP Problem and the MFV
  Principle}},  Phys. Lett. B {\bf 684} (2010) 147--153,
  [\href{http://arxiv.org/abs/0906.4551}{{\tt arXiv:0906.4551}}].

\bibitem{Grinstein:2010ve}
B.~Grinstein, M.~Redi, and G.~Villadoro, {\it {Low Scale Flavor Gauge
  Symmetries}},  JHEP {\bf 11} (2010) 067,
  [\href{http://arxiv.org/abs/1009.2049}{{\tt arXiv:1009.2049}}].

\bibitem{Feldmann:2010yp}
T.~Feldmann, {\it {See-Saw Masses for Quarks and Leptons in SU(5)}},  JHEP {\bf
  04} (2011) 043, [\href{http://arxiv.org/abs/1010.2116}{{\tt
  arXiv:1010.2116}}].

\bibitem{Guadagnoli:2011id}
D.~Guadagnoli, R.~N. Mohapatra, and I.~Sung, {\it {Gauged Flavor Group with
  Left-Right Symmetry}},  JHEP {\bf 04} (2011) 093,
  [\href{http://arxiv.org/abs/1103.4170}{{\tt arXiv:1103.4170}}].

\bibitem{Buras:2011zb}
A.~J. Buras, L.~Merlo, and E.~Stamou, {\it {The Impact of Flavour Changing
  Neutral Gauge Bosons on $\bar{B} -> X_s \gamma$}},  JHEP {\bf 08} (2011) 124,
  [\href{http://arxiv.org/abs/1105.5146}{{\tt arXiv:1105.5146}}].

\bibitem{Buras:2011wi}
A.~J. Buras, M.~V. Carlucci, L.~Merlo, and E.~Stamou, {\it {Phenomenology of a
  Gauged $SU(3)^3$ Flavour Model}},  JHEP {\bf 03} (2012) 088,
  [\href{http://arxiv.org/abs/1112.4477}{{\tt arXiv:1112.4477}}].

\bibitem{Alonso:2012jc}
R.~Alonso, M.~B. Gavela, L.~Merlo, S.~Rigolin, and J.~Yepes, {\it {Minimal
  Flavour Violation with Strong Higgs Dynamics}},  JHEP {\bf 06} (2012) 076,
  [\href{http://arxiv.org/abs/1201.1511}{{\tt arXiv:1201.1511}}].

\bibitem{Alonso:2012pz}
R.~Alonso, M.~B. Gavela, L.~Merlo, S.~Rigolin, and J.~Yepes, {\it {Flavor with
  a light dynamical ''Higgs particle''}},  Phys. Rev. D {\bf 87} (2013), no.~5
  055019, [\href{http://arxiv.org/abs/1212.3307}{{\tt arXiv:1212.3307}}].

\bibitem{Lopez-Honorez:2013wla}
L.~Lopez-Honorez and L.~Merlo, {\it {Dark matter within the minimal flavour
  violation ansatz}},  Phys. Lett. B {\bf 722} (2013) 135--143,
  [\href{http://arxiv.org/abs/1303.1087}{{\tt arXiv:1303.1087}}].

\bibitem{Barbieri:2014tja}
R.~Barbieri, D.~Buttazzo, F.~Sala, and D.~M. Straub, {\it {Flavour physics and
  flavour symmetries after the first LHC phase}},  JHEP {\bf 05} (2014) 105,
  [\href{http://arxiv.org/abs/1402.6677}{{\tt arXiv:1402.6677}}].

\bibitem{Alonso:2016onw}
R.~Alonso, E.~Fernandez~Martinez, M.~B. Gavela, B.~Grinstein, L.~Merlo, and
  P.~Quilez, {\it {Gauged Lepton Flavour}},  JHEP {\bf 12} (2016) 119,
  [\href{http://arxiv.org/abs/1609.05902}{{\tt arXiv:1609.05902}}].

\bibitem{Crivellin:2016ejn}
A.~Crivellin, J.~Fuentes-Martin, A.~Greljo, and G.~Isidori, {\it {Lepton Flavor
  Non-Universality in B decays from Dynamical Yukawas}},  Phys. Lett. B {\bf
  766} (2017) 77--85, [\href{http://arxiv.org/abs/1611.02703}{{\tt
  arXiv:1611.02703}}].

\bibitem{Dinh:2017smk}
D.~N. Dinh, L.~Merlo, S.~T. Petcov, and R.~Vega-\'Alvarez, {\it {Revisiting
  Minimal Lepton Flavour Violation in the Light of Leptonic CP Violation}},
  JHEP {\bf 07} (2017) 089, [\href{http://arxiv.org/abs/1705.09284}{{\tt
  arXiv:1705.09284}}].

\bibitem{Merlo:2018rin}
L.~Merlo and S.~Rosauro-Alcaraz, {\it {Predictive Leptogenesis from Minimal
  Lepton Flavour Violation}},  JHEP {\bf 07} (2018) 036,
  [\href{http://arxiv.org/abs/1801.03937}{{\tt arXiv:1801.03937}}].

\bibitem{Arias-Aragon:2020qip}
F.~Arias-Aragon, E.~Fernandez-Martinez, M.~Gonzalez-Lopez, and L.~Merlo, {\it
  {Neutrino Masses and Hubble Tension via a Majoron in MFV}},  Eur. Phys. J. C
  {\bf 81} (2021), no.~1 28, [\href{http://arxiv.org/abs/2009.01848}{{\tt
  arXiv:2009.01848}}].

\bibitem{Alonso-Gonzalez:2021jsa}
J.~Alonso-Gonz\'alez, L.~Merlo, and S.~Pokorski, {\it {A new bound on CP
  violation in the \ensuremath{\tau} lepton Yukawa coupling and electroweak
  baryogenesis}},  JHEP {\bf 06} (2021) 166,
  [\href{http://arxiv.org/abs/2103.16569}{{\tt arXiv:2103.16569}}].

\bibitem{Alonso-Gonzalez:2021tpo}
J.~Alonso-Gonzalez, A.~de~Giorgi, L.~Merlo, and S.~Pokorski, {\it {Searching
  for BSM Physics in Yukawa Couplings and Flavour Symmetries}},
  \href{http://arxiv.org/abs/2109.07490}{{\tt arXiv:2109.07490}}.

\bibitem{Feldmann:2009dc}
T.~Feldmann, M.~Jung, and T.~Mannel, {\it {Sequential Flavour Symmetry
  Breaking}},  Phys. Rev. D {\bf 80} (2009) 033003,
  [\href{http://arxiv.org/abs/0906.1523}{{\tt arXiv:0906.1523}}].

\bibitem{Alonso:2011yg}
R.~Alonso, M.~B. Gavela, L.~Merlo, and S.~Rigolin, {\it {On the scalar
  potential of minimal flavour violation}},  JHEP {\bf 07} (2011) 012,
  [\href{http://arxiv.org/abs/1103.2915}{{\tt arXiv:1103.2915}}].

\bibitem{Nardi:2011st}
E.~Nardi, {\it {Naturally large Yukawa hierarchies}},  Phys. Rev. D {\bf 84}
  (2011) 036008, [\href{http://arxiv.org/abs/1105.1770}{{\tt
  arXiv:1105.1770}}].

\bibitem{Alonso:2012fy}
R.~Alonso, M.~B. Gavela, D.~Hernandez, and L.~Merlo, {\it {On the Potential of
  Leptonic Minimal Flavour Violation}},  Phys. Lett. B {\bf 715} (2012)
  194--198, [\href{http://arxiv.org/abs/1206.3167}{{\tt arXiv:1206.3167}}].

\bibitem{Alonso:2013mca}
R.~Alonso, M.~B. Gavela, D.~Hern\'andez, L.~Merlo, and S.~Rigolin, {\it
  {Leptonic Dynamical Yukawa Couplings}},  JHEP {\bf 08} (2013) 069,
  [\href{http://arxiv.org/abs/1306.5922}{{\tt arXiv:1306.5922}}].

\bibitem{Alonso:2013nca}
R.~Alonso, M.~B. Gavela, G.~Isidori, and L.~Maiani, {\it {Neutrino Mixing and
  Masses from a Minimum Principle}},  JHEP {\bf 11} (2013) 187,
  [\href{http://arxiv.org/abs/1306.5927}{{\tt arXiv:1306.5927}}].

\bibitem{Arias-Aragon:2020bzy}
F.~Arias-Arag\'on, C.~Bouthelier-Madre, J.~M. Cano, and L.~Merlo, {\it {Data
  Driven Flavour Model}},  Eur. Phys. J. C {\bf 80} (2020), no.~9 854,
  [\href{http://arxiv.org/abs/2003.05941}{{\tt arXiv:2003.05941}}].

\bibitem{Dolan:2018yqy}
M.~J. Dolan, T.~P. Dutka, and R.~R. Volkas, {\it {Low-scale leptogenesis with
  minimal lepton flavor violation}},  Phys. Rev. D {\bf 99} (2019), no.~12
  123508, [\href{http://arxiv.org/abs/1812.11964}{{\tt arXiv:1812.11964}}].

\bibitem{Ma:2009gu}
E.~Ma, {\it {Radiative inverse seesaw mechanism for nonzero neutrino mass}},
  Phys. Rev. D {\bf 80} (2009) 013013,
  [\href{http://arxiv.org/abs/0904.4450}{{\tt arXiv:0904.4450}}].

\bibitem{Bazzocchi:2010dt}
F.~Bazzocchi, {\it {Minimal Dynamical Inverse See Saw}},  Phys. Rev. D {\bf 83}
  (2011) 093009, [\href{http://arxiv.org/abs/1011.6299}{{\tt
  arXiv:1011.6299}}].

\bibitem{DeRomeri:2017oxa}
V.~De~Romeri, E.~Fernandez-Martinez, J.~Gehrlein, P.~A.~N. Machado, and
  V.~Niro, {\it {Dark Matter and the elusive Z' in a dynamical Inverse Seesaw
  scenario}},  JHEP {\bf 10} (2017) 169,
  [\href{http://arxiv.org/abs/1707.08606}{{\tt arXiv:1707.08606}}].

\bibitem{Mandal:2021acg}
S.~Mandal, J.~C. Rom\~ao, R.~Srivastava, and J.~W.~F. Valle, {\it {Dynamical
  inverse seesaw mechanism as a simple benchmark for electroweak breaking and
  Higgs boson studies}},  JHEP {\bf 07} (2021) 029,
  [\href{http://arxiv.org/abs/2103.02670}{{\tt arXiv:2103.02670}}].

\bibitem{Fernandez-Martinez:2021ypo}
E.~Fernandez-Martinez, M.~Pierre, E.~Pinsard, and S.~Rosauro-Alcaraz, {\it
  {Inverse Seesaw, dark matter and the Hubble tension}},  Eur. Phys. J. C {\bf
  81} (2021), no.~10 954, [\href{http://arxiv.org/abs/2106.05298}{{\tt
  arXiv:2106.05298}}].

\bibitem{Peccei:1977hh}
R.~D. Peccei and H.~R. Quinn, {\it {CP Conservation in the Presence of
  Instantons}},  Phys. Rev. Lett. {\bf 38} (1977) 1440--1443.

\bibitem{Wilczek:1977pj}
F.~Wilczek, {\it {Problem of Strong $P$ and $T$ Invariance in the Presence of
  Instantons}},  Phys. Rev. Lett. {\bf 40} (1978) 279--282.

\bibitem{Weinberg:1977ma}
S.~Weinberg, {\it {A New Light Boson?}},  Phys. Rev. Lett. {\bf 40} (1978)
  223--226.

\bibitem{Arias-Aragon:2017eww}
F.~Arias-Aragon and L.~Merlo, {\it {The Minimal Flavour Violating Axion}},
  JHEP {\bf 10} (2017) 168, [\href{http://arxiv.org/abs/1709.07039}{{\tt
  arXiv:1709.07039}}]. [Erratum: JHEP 11, 152 (2019)].

\bibitem{CDF:2022hxs}
{\bf CDF} Collaboration, T.~Aaltonen {\em et.~al.}, {\it {High-precision
  measurement of the $W$ boson mass with the CDF II detector}},  Science {\bf
  376} (2022), no.~6589 170--176.

\bibitem{Bertuzzo:2009im}
E.~Bertuzzo, P.~Di~Bari, F.~Feruglio, and E.~Nardi, {\it {Flavor symmetries,
  leptogenesis and the absolute neutrino mass scale}},  JHEP {\bf 11} (2009)
  036, [\href{http://arxiv.org/abs/0908.0161}{{\tt arXiv:0908.0161}}].

\bibitem{AristizabalSierra:2009ex}
D.~Aristizabal~Sierra, F.~Bazzocchi, I.~de~Medeiros~Varzielas, L.~Merlo, and
  S.~Morisi, {\it {Tri-Bimaximal Lepton Mixing and Leptogenesis}},  Nucl. Phys.
  B {\bf 827} (2010) 34--58, [\href{http://arxiv.org/abs/0908.0907}{{\tt
  arXiv:0908.0907}}].

\bibitem{Froggatt:1978nt}
C.~D. Froggatt and H.~B. Nielsen, {\it {Hierarchy of Quark Masses, Cabibbo
  Angles and CP Violation}},  Nucl. Phys. B {\bf 147} (1979) 277--298.

\bibitem{Raffelt:2006cw}
G.~G. Raffelt, {\it {Astrophysical axion bounds}},  Lect. Notes Phys. {\bf 741}
  (2008) 51--71, [\href{http://arxiv.org/abs/hep-ph/0611350}{{\tt
  hep-ph/0611350}}].

\bibitem{Mimasu:2014nea}
K.~Mimasu and V.~Sanz, {\it {ALPs at Colliders}},  JHEP {\bf 06} (2015) 173,
  [\href{http://arxiv.org/abs/1409.4792}{{\tt arXiv:1409.4792}}].

\bibitem{Vinyoles:2015aba}
N.~Vinyoles, A.~Serenelli, F.~L. Villante, S.~Basu, J.~Redondo, and J.~Isern,
  {\it {New axion and hidden photon constraints from a solar data global fit}},
   JCAP {\bf 10} (2015) 015, [\href{http://arxiv.org/abs/1501.01639}{{\tt
  arXiv:1501.01639}}].

\bibitem{DiLuzio:2016sbl}
L.~Di~Luzio, F.~Mescia, and E.~Nardi, {\it {Redefining the Axion Window}},
  Phys. Rev. Lett. {\bf 118} (2017), no.~3 031801,
  [\href{http://arxiv.org/abs/1610.07593}{{\tt arXiv:1610.07593}}].

\bibitem{Brivio:2017ije}
I.~Brivio, M.~B. Gavela, L.~Merlo, K.~Mimasu, J.~M. No, R.~del Rey, and
  V.~Sanz, {\it {ALPs Effective Field Theory and Collider Signatures}},  Eur.
  Phys. J. C {\bf 77} (2017), no.~8 572,
  [\href{http://arxiv.org/abs/1701.05379}{{\tt arXiv:1701.05379}}].

\bibitem{Dolan:2017osp}
M.~J. Dolan, T.~Ferber, C.~Hearty, F.~Kahlhoefer, and K.~Schmidt-Hoberg, {\it
  {Revised constraints and Belle II sensitivity for visible and invisible
  axion-like particles}},  JHEP {\bf 12} (2017) 094,
  [\href{http://arxiv.org/abs/1709.00009}{{\tt arXiv:1709.00009}}]. [Erratum:
  JHEP 03, 190 (2021)].

\bibitem{Bauer:2017ris}
M.~Bauer, M.~Neubert, and A.~Thamm, {\it {Collider Probes of Axion-Like
  Particles}},  JHEP {\bf 12} (2017) 044,
  [\href{http://arxiv.org/abs/1708.00443}{{\tt arXiv:1708.00443}}].

\bibitem{Alonso-Alvarez:2018irt}
G.~Alonso-\'Alvarez, M.~B. Gavela, and P.~Quilez, {\it {Axion couplings to
  electroweak gauge bosons}},  Eur. Phys. J. C {\bf 79} (2019), no.~3 223,
  [\href{http://arxiv.org/abs/1811.05466}{{\tt arXiv:1811.05466}}].

\bibitem{Bauer:2018uxu}
M.~Bauer, M.~Heiles, M.~Neubert, and A.~Thamm, {\it {Axion-Like Particles at
  Future Colliders}},  Eur. Phys. J. C {\bf 79} (2019), no.~1 74,
  [\href{http://arxiv.org/abs/1808.10323}{{\tt arXiv:1808.10323}}].

\bibitem{Gavela:2019wzg}
M.~B. Gavela, R.~Houtz, P.~Quilez, R.~Del~Rey, and O.~Sumensari, {\it {Flavor
  constraints on electroweak ALP couplings}},  Eur. Phys. J. C {\bf 79} (2019),
  no.~5 369, [\href{http://arxiv.org/abs/1901.02031}{{\tt arXiv:1901.02031}}].

\bibitem{Merlo:2019anv}
L.~Merlo, F.~Pobbe, S.~Rigolin, and O.~Sumensari, {\it {Revisiting the
  production of ALPs at B-factories}},  JHEP {\bf 06} (2019) 091,
  [\href{http://arxiv.org/abs/1905.03259}{{\tt arXiv:1905.03259}}].

\bibitem{Arias-Aragon:2020qtn}
F.~Arias-Arag\'on, F.~D'eramo, R.~Z. Ferreira, L.~Merlo, and A.~Notari, {\it
  {Cosmic Imprints of XENON1T Axions}},  JCAP {\bf 11} (2020) 025,
  [\href{http://arxiv.org/abs/2007.06579}{{\tt arXiv:2007.06579}}].

\bibitem{Arias-Aragon:2020shv}
F.~Arias-Arag\'on, F.~D'Eramo, R.~Z. Ferreira, L.~Merlo, and A.~Notari, {\it
  {Production of Thermal Axions across the ElectroWeak Phase Transition}},
  JCAP {\bf 03} (2021) 090, [\href{http://arxiv.org/abs/2012.04736}{{\tt
  arXiv:2012.04736}}].

\bibitem{Viaux:2013lha}
N.~Viaux, M.~Catelan, P.~B. Stetson, G.~Raffelt, J.~Redondo, A.~A.~R. Valcarce,
  and A.~Weiss, {\it {Neutrino and axion bounds from the globular cluster M5
  (NGC 5904)}},  Phys. Rev. Lett. {\bf 111} (2013) 231301,
  [\href{http://arxiv.org/abs/1311.1669}{{\tt arXiv:1311.1669}}].

\bibitem{Preskill:1982cy}
J.~Preskill, M.~B. Wise, and F.~Wilczek, {\it {Cosmology of the Invisible
  Axion}},  Phys. Lett. B {\bf 120} (1983) 127--132.

\bibitem{Abbott:1982af}
L.~F. Abbott and P.~Sikivie, {\it {A Cosmological Bound on the Invisible
  Axion}},  Phys. Lett. B {\bf 120} (1983) 133--136.

\bibitem{Dine:1982ah}
M.~Dine and W.~Fischler, {\it {The Not So Harmless Axion}},  Phys. Lett. B {\bf
  120} (1983) 137--141.

\bibitem{Gorghetto:2018myk}
M.~Gorghetto, E.~Hardy, and G.~Villadoro, {\it {Axions from Strings: the
  Attractive Solution}},  JHEP {\bf 07} (2018) 151,
  [\href{http://arxiv.org/abs/1806.04677}{{\tt arXiv:1806.04677}}].

\bibitem{Gorghetto:2020qws}
M.~Gorghetto, E.~Hardy, and G.~Villadoro, {\it {More axions from strings}},
  SciPost Phys. {\bf 10} (2021), no.~2 050,
  [\href{http://arxiv.org/abs/2007.04990}{{\tt arXiv:2007.04990}}].

\bibitem{Antusch:2014woa}
S.~Antusch and O.~Fischer, {\it {Non-unitarity of the leptonic mixing matrix:
  Present bounds and future sensitivities}},  JHEP {\bf 10} (2014) 094,
  [\href{http://arxiv.org/abs/1407.6607}{{\tt arXiv:1407.6607}}].

\bibitem{Fernandez-Martinez:2016lgt}
E.~Fernandez-Martinez, J.~Hernandez-Garcia, and J.~Lopez-Pavon, {\it {Global
  constraints on heavy neutrino mixing}},  JHEP {\bf 08} (2016) 033,
  [\href{http://arxiv.org/abs/1605.08774}{{\tt arXiv:1605.08774}}].

\bibitem{deGouvea:2008nm}
A.~de~Gouvea and J.~Jenkins, {\it {The Physical Range of Majorana Neutrino
  Mixing Parameters}},  Phys. Rev. D {\bf 78} (2008) 053003,
  [\href{http://arxiv.org/abs/0804.3627}{{\tt arXiv:0804.3627}}].

\bibitem{Casas:2001sr}
J.~A. Casas and A.~Ibarra, {\it {Oscillating neutrinos and $\mu \to e,
  \gamma$}},  Nucl. Phys. B {\bf 618} (2001) 171--204,
  [\href{http://arxiv.org/abs/hep-ph/0103065}{{\tt hep-ph/0103065}}].

\bibitem{Cirigliano:2006nu}
V.~Cirigliano, G.~Isidori, and V.~Porretti, {\it {CP violation and Leptogenesis
  in models with Minimal Lepton Flavour Violation}},  Nucl. Phys. B {\bf 763}
  (2007) 228--246, [\href{http://arxiv.org/abs/hep-ph/0607068}{{\tt
  hep-ph/0607068}}].

\bibitem{Fernandez-Martinez:2007iaa}
E.~Fernandez-Martinez, M.~B. Gavela, J.~Lopez-Pavon, and O.~Yasuda, {\it
  {CP-violation from non-unitary leptonic mixing}},  Phys. Lett. B {\bf 649}
  (2007) 427--435, [\href{http://arxiv.org/abs/hep-ph/0703098}{{\tt
  hep-ph/0703098}}].

\bibitem{Langacker:1988ur}
P.~Langacker and D.~London, {\it {Mixing Between Ordinary and Exotic
  Fermions}},  Phys. Rev. D {\bf 38} (1988) 886.

\bibitem{Antusch:2006vwa}
S.~Antusch, C.~Biggio, E.~Fernandez-Martinez, M.~B. Gavela, and J.~Lopez-Pavon,
  {\it {Unitarity of the Leptonic Mixing Matrix}},  JHEP {\bf 10} (2006) 084,
  [\href{http://arxiv.org/abs/hep-ph/0607020}{{\tt hep-ph/0607020}}].

\bibitem{ParticleDataGroup:2020ssz}
{\bf Particle Data Group} Collaboration, P.~A. Zyla {\em et.~al.}, {\it {Review
  of Particle Physics}},  PTEP {\bf 2020} (2020), no.~8 083C01.

\bibitem{Fernandez-Martinez:2015hxa}
E.~Fernandez-Martinez, J.~Hernandez-Garcia, J.~Lopez-Pavon, and M.~Lucente,
  {\it {Loop level constraints on Seesaw neutrino mixing}},  JHEP {\bf 10}
  (2015) 130, [\href{http://arxiv.org/abs/1508.03051}{{\tt arXiv:1508.03051}}].

\bibitem{Janot:2019oyi}
P.~Janot and S.~Jadach, {\it {Improved Bhabha cross section at LEP and the
  number of light neutrino species}},  Phys. Lett. B {\bf 803} (2020) 135319,
  [\href{http://arxiv.org/abs/1912.02067}{{\tt arXiv:1912.02067}}].

\bibitem{Alonso:2012ji}
R.~Alonso, M.~Dhen, M.~B. Gavela, and T.~Hambye, {\it {Muon conversion to
  electron in nuclei in type-I seesaw models}},  JHEP {\bf 01} (2013) 118,
  [\href{http://arxiv.org/abs/1209.2679}{{\tt arXiv:1209.2679}}].

\bibitem{Kitano:2002mt}
R.~Kitano, M.~Koike, and Y.~Okada, {\it {Detailed calculation of lepton flavor
  violating muon electron conversion rate for various nuclei}},  Phys. Rev. D
  {\bf 66} (2002) 096002, [\href{http://arxiv.org/abs/hep-ph/0203110}{{\tt
  hep-ph/0203110}}]. [Erratum: Phys.Rev.D 76, 059902 (2007)].

\bibitem{Suzuki:1987jf}
T.~Suzuki, D.~F. Measday, and J.~P. Roalsvig, {\it {Total Nuclear Capture Rates
  for Negative Muons}},  Phys. Rev. C {\bf 35} (1987) 2212.

\bibitem{Esteban:2020cvm}
I.~Esteban, M.~C. Gonzalez-Garcia, M.~Maltoni, T.~Schwetz, and A.~Zhou, {\it
  {The fate of hints: updated global analysis of three-flavor neutrino
  oscillations}},  JHEP {\bf 09} (2020) 178,
  [\href{http://arxiv.org/abs/2007.14792}{{\tt arXiv:2007.14792}}].

\bibitem{Planck:2018vyg}
{\bf Planck} Collaboration, N.~Aghanim {\em et.~al.}, {\it {Planck 2018
  results. VI. Cosmological parameters}},  Astron. Astrophys. {\bf 641} (2020)
  A6, [\href{http://arxiv.org/abs/1807.06209}{{\tt arXiv:1807.06209}}].
  [Erratum: Astron.Astrophys. 652, C4 (2021)].

\bibitem{Baldini:2021kfb}
A.~M. Baldini and T.~Mori, {\it {MEG: Muon to Electron and Gamma}},  SciPost
  Phys. Proc. {\bf 5} (2021) 019.

\bibitem{SnowmassCLFVTau}
S.~Banerjee, V.~Cirigliano, M.~Dam, A.~Deshpande, L.~Fiorini, K.~Fuyuto,
  C.~Gal, T.~Husek, E.~Mereghetti, K.~Monsálvez-Pozo, H.~Peng, F.~Polci,
  J.~Portolés, A.~Rostomyan, M.~H. Villanueva, B.~Yan, J.~Zhang, and X.~Zhou,
  {\it Snowmass 2021 white paper: Charged lepton flavor violation in the tau
  sector},  2022.

\bibitem{Kutschke:2011ux}
R.~K. Kutschke, {\it {The Mu2e Experiment at Fermilab}},  in {\em {31st
  International Symposium on Physics In Collision}}, 12, 2011.
\newblock \href{http://arxiv.org/abs/1112.0242}{{\tt arXiv:1112.0242}}.

\bibitem{Barlow:2011zza}
R.~J. Barlow, {\it {The PRISM/PRIME project}},  Nucl. Phys. B Proc. Suppl. {\bf
  218} (2011) 44--49.

\bibitem{Abdullahi:2022jlv}
A.~M. Abdullahi {\em et.~al.}, {\it {The Present and Future Status of Heavy
  Neutral Leptons}},  in {\em {2022 Snowmass Summer Study}}, 3, 2022.
\newblock \href{http://arxiv.org/abs/2203.08039}{{\tt arXiv:2203.08039}}.

\bibitem{Bryman:2021teu}
D.~Bryman, V.~Cirigliano, A.~Crivellin, and G.~Inguglia, {\it {Testing Lepton
  Flavor Universality with Pion, Kaon, Tau, and Beta Decays}},
  \href{http://arxiv.org/abs/2111.05338}{{\tt arXiv:2111.05338}}.

\end{thebibliography}\endgroup
\bibliographystyle{BiblioStyle}

\end{document}